\documentclass[11pt,english,letterpaper]{elsarticle}
\usepackage{mathptmx}
\usepackage[T1]{fontenc}
\usepackage[latin9]{inputenc}
\usepackage{geometry}
\usepackage{amsbsy}
\usepackage{amstext}
\usepackage{amssymb}
\usepackage{graphicx}

\makeatletter

\providecommand{\tabularnewline}{\\}

\usepackage{fancybox}

\newcommand{\myBox}[1]{%
  \fbox{$ \displaystyle #1 $}%
}

\usepackage{epsfig}

\makeatother

\usepackage{babel}
\begin{document}

\title{{\Large{}A scalable multidimensional fully implicit solver for Hall
magnetohydrodynamics}}

\author[lanl]{L. Chacón\corref{cor1}}

\ead{chacon@lanl.gov}

\cortext[cor1]{Corresponding author}

\address[lanl]{Los Alamos National Laboratory, Los Alamos, NM 87545, USA}
\begin{abstract}
We propose an optimally performant fully implicit algorithm for the
Hall magnetohydrodynamics (HMHD) equations based on multigrid-preconditioned
Jacobian-free Newton-Krylov methods. HMHD is a challenging system
to solve numerically because it supports stiff fast dispersive waves.
The preconditioner is formulated using an operator-split approximate
block factorization (Schur complement), informed by physics insight.
We use a vector-potential formulation (instead of a magnetic field
one) allow a clean segregation of the problematic $\nabla\times\nabla\times$
operator in the electron Ohm's law subsystem. This segregation allows
the formulation of an effective damped block-Jacobi smoother for multigrid.
We demonstrate by analysis that our proposed block-Jacobi iteration
is convergent and has the smoothing property. The resulting HMHD solver
is verified linearly with wave propagation examples, and nonlinearly
with the GEM challenge reconnection problem by comparison against
another HMHD code. We demonstrate the excellent algorithmic and parallel
performance of the algorithm up to 16384 MPI tasks in two dimensions.
\end{abstract}
\maketitle

\section{Introduction}

We propose a scalable implicit algorithm for the Hall magnetohydrodynamics
(HMHD) model, which is of relevance in the modeling of astrophysical,
magnetospheric, and laboratory (e.g., fusion) plasmas. HMHD is well
known to support fast dispersive waves, which become very stiff numerically
with increasing spatial resolution, resulting in explicit Courant-Friedrichs-Lewy
(CFL) timestep stability constraints scaling quadratically with the
mesh spacing rather than linearly. As a result, the explicit integration
of this system becomes very expensive for sufficiently refined meshes,
encouraging the exploration of implicit methods.

Several authors have attempted to build implicitness in the solution
algorithm of the HMHD equations over the years. The earliest study
we are aware of is Ref. \citep{harned-mikic-si}, where a semi-implicit
solver for the HMHD system based on a pseudospectral discretization
in cylindrical geometry was proposed for out-of-plane uniform magnetic
fields (which incidentally do not propagate the problematic stiff
hyperbolic waves mentioned earlier). The stability analysis of the
semi-implicit time-advance formulation in the general case (including
stiff hyperbolic waves) was studied in Ref. \citep{sovinec-jcp-10_hall-mhd},
concluding that strict time-centering was needed for numerical stability.
Several authors have explored fully implicit HMHD formulations, either
in its reduced form (i.e., assuming plasma incompressibility) \citep{chacon-JCP-hall,jardin-pop-05-mhd,kai-mrc-06,dobrian-jcp-07-asm_mhd,chacon2016scalable}
or in primitive-variable form \citep{kai-mrc-06,chacon-jpcs-08-scidac,arnold2008semi,toth-jcp-08-hall_amr,lutjens2010xtor,laakmann2023structure}.
Of these, many of them employ either direct solvers \citep{jardin-pop-05-mhd,kai-mrc-06,dobrian-jcp-07-asm_mhd,lutjens2010xtor,laakmann2023structure}
or incomplete LU factorizations \citep{toth-jcp-08-hall_amr}, which
are known not to scale well either algorithmically under grid refinement,
or in parallel (and in fact little information is provided about either
scaling in the references). Scalable fully implicit solutions for
the reduced HMHD system \citep{chacon-JCP-hall,chacon2016scalable}
and the primitive-variable HMHD system in a non-standard ion formulation
\citep{chacon-jpcs-08-scidac} were proposed that show significant
promise. However, the latter formulation (relevant to the present
study) has been found by the author to be prone to numerical noise
from fast dispersive waves, and problematic to solve when attempting
to include high-order dissipation (via hyperresistivity, mimicking
electron viscosity \citep{gem}) to control it. Other authors have
attempted to bypass the numerical stiffness challenges of HMHD either
by adding a fictitious displacement current term in Ampere's law \citep{zhao2014positivity,thoma2021implicit}
or by reverting back to a multi-fluid formulation with promising approximate
block-factorization preconditioners \citep{phillips2020enabling,crockatt2022implicit}.
The former approach is problematic because the formulation is physically
inconsistent, and can be shown to introduce instability \citep{thoma2021implicit}.
The latter introduces a large number of additional equations, and
features fast charge-separation and lightwave timescales that complicate
an implicit treatment (although these could be important in some physical
contexts such as the treatment of boundary sheaths).

This paper builds on earlier work \citep{chacon-pop-08-3dmhd} where
a scalable, parallel, fully implicit, fully nonlinear solver for the
3D resistive compressible MHD equations was demonstrated, and augments
it to include electron Hall and pressure effects. As in the reference,
we base our nonlinear solver approach on the Newton-Raphson iterative
algorithm. Krylov iterative techniques \citep{saad-book}, implemented
Jacobian-free \citep{chan-siam-84-nk,saad-nlkrylov} (i.e., without
ever forming and storing the Jacobian matrix) for memory efficiency,
are employed for the required algebraic matrix inversions. Here, FGMRES
(Flexible Generalized Minimal RESiduals \citep{saad-gmres,saad-fgmres})
is employed as the Krylov solver of choice, due to the lack of symmetry
in the algebraic system of interest. The flexible character of FGMRES
allows the preconditioner to change between successive GMRES iterations.
We will exploit this feature in our implementation.

The efficiency and scalability of Krylov methods depends strongly
on adequate preconditioning \citep{saad-book}. In Refs. \citep{chacon-JCP-hall,chacon-pop-08-3dmhd},
a suitable multigrid-based preconditioner is developed for the MHD
system around the parabolization concept, whereby a hyperbolic system
is reformulated as a parabolic one using a Schur factorization, which
is in turn amenable to a multilevel treatment. The parabolization
procedure is generalized here to HMHD. Using physics insight, we approximate
the exact Schur factorization with an operator-split procedure, where
the stiff electron subsystem (which we will term electron MHD, or
EMHD) is inverted first, followed by the ion MHD subsystem inversion
proposed in Ref. \citep{chacon-pop-08-3dmhd}. All subsystems are
solved using a fully parallel geometric multigrid (MG) implementation
based on a coarse-grid parallel-domain agglomeration strategy (similar
to that proposed in Refs. \citep{hulsemann2006parallel,petsc-mg-parallel-2016}).
For the EMHD subsystem (which is the stiffest one in HMHD), we propose
a novel six-equation block damped-Jacobi smoothing strategy that is
shown by analysis to lead to convergent Jacobi iterations and to possess
the smoothing property. Damped Jacobi is demonstrated in Ref. \citep{arnold2008semi}
to be a viable smoother for a semi-implicit discretization of the
2D EMHD system (without hyperresistivity), and is implemented in a
nonlinear MG solver. However, no grid-convergence study is offered
in the reference, and relatively small implicit timesteps (of $\mathcal{O}(40)$
explicit CFLs) are used. Here, we extend the analysis to full 3D geometry
(which requires dealing with the $\nabla\times\nabla\times$ operator,
very challenging for geometric MG solvers), include hyperresistivity,
and push implicit timesteps to $8400$ CFLs without loss of algorithmic
performance. The formulation is enabled by a vector-potential formulation
of the HMHD system (instead of the more common magnetic-field one).
The resulting smoother is able to deal seamlessly with both the whistler-wave
and the fourth-order hyper-resistivity operators, both featuring $\nabla\times\nabla\times$
terms. As a result of these advances, the preconditioner renders the
Jacobian-free Newton-Krylov (JFNK) solver optimally scalable algorithmically
and in parallel (in a weak scaling sense) up to the largest number
of processors considered in this study (16384 processors).

The rest of the paper is organized as follows. Section \ref{sec:model}
introduces the base model equations. Sec. \ref{sec:preconditioning}
discusses the details of the proposed physics-based preconditioner.
Numerical details of our implementation, including the multigrid (MG)
treatment of the hyperbolically stiff HMHD system, are provided in
Sec. \ref{sec:Numerical-details} (with detailed analysis of the smoothing
properties of our block-Jacobi treatment of the Hall MHD equations
provided in \ref{sec:JB_analysis}). Linear and nonlinear verification
results of our implementation using wave propagation examples and
the GEM challenge problem \citep{gem}, respectively, along with a
weak parallel scalability study of our implicit solver up to 16384
cores, are presented in Sec. \ref{sec:Numerical-results}. Finally,
we conclude in Sec. \ref{sec:conclusions}.

\section{Vector-potential formulation of the Hall MHD model}

\label{sec:model}

In Alfvénic units (i.e, Alfvén speed $v_{A}=B_{0}/\sqrt{\mu_{0}\rho_{0}}$,
Alfvén time $\tau_{A}=L/v_{A}$, where $B_{0}$, $\rho_{0}$, and
$L$ are the reference magnetic field, density, and length, respectively),
the HMHD model of interest for this study reads:
\begin{eqnarray}
\frac{\partial\rho}{\partial t} & + & \nabla\cdot(\rho\mathbf{v})=0,\label{eq:continuity-dimlss}\\
\frac{\partial\mathbf{B}}{\partial t} & + & \nabla\times\mathbf{E}=0\label{eq:faraday}\\
\frac{\partial(\rho\mathbf{v})}{\partial t}+\nabla\cdot\left[\rho\mathbf{\mathbf{v}v}-\mathbf{B}\mathbf{B}\right. & + & \left.\mathbb{I}(p+\frac{B^{2}}{2})+\Pi_{i}\right]=0,\label{eq:EOM-dimlss}\\
\frac{\partial p}{\partial t}+\nabla\cdot(\mathbf{v}^{*}p) & + & (\gamma-1)p\nabla\cdot\mathbf{v}^{*}=(\gamma-1)(Q-\nabla\cdot\mathbf{q}),\label{eq:temperature-dimlss}
\end{eqnarray}
with $\mathbb{I}$ the identity operator, $\rho$ the particle density,
$\mathbf{v}$ the plasma velocity, $\mathbf{B}$ the magnetic field,
$\mathbf{E}$ the electric field, $p=\rho T=(1+\alpha)\rho T_{e}$
the total pressure, and $\mathbf{v}^{*}=\frac{\alpha\mathbf{v}+\mathbf{v}_{e}}{\alpha+1}=\mathbf{v}-\frac{d_{i}}{1+\alpha}\frac{\mathbf{j}}{\rho}$.
Here, $\alpha=T_{i}/T_{e}$ is a fixed ion-to-electron temperature
ratio, $\mathbf{v}_{e}=\mathbf{v}-d_{i}\,\mathbf{j}/\rho$ is the
electron velocity, $\mathbf{j}=\nabla\times\mathbf{B}$ is the current,
and $d_{i}=c/\omega_{pi}L$ the dimensionless ion inertial length.
The velocity $\mathbf{v}^{*}$ in the pressure equation, Eq. \ref{eq:temperature-dimlss},
is derived by adding the ion and electron internal energy evolution
equations \citep{braginskii}, and using the fixed-temperature-ratio
assumption to ensure existence of a local energy conservation law
for fixed temperature ratio $\alpha$, as we outline below. Note that
the pressure equation reverts to the electron pressure equation $p_{e}=p/(1+\alpha)$
for cold ions ($\alpha\ll1$), and to the ion pressure equation $p_{i}=p\alpha/(1+\alpha)$
for cold electrons ($\alpha\gg1$).

In these equations, $\eta$ is the normalized resistivity (inverse
Lundquist number), $\mathbf{q}$ is the heat flux, and $Q=\eta j^{2}-\sum_{s=i,e}\Pi_{s}:\nabla\mathbf{v}_{s}$
contains the joule and viscous heating sources. Simple closures for
the heat flux $\mathbf{q}=-\kappa\nabla T_{e}$ and the viscous stress
tensors $\Pi_{s}=-\rho\nu_{s}\nabla\mathbf{v}_{s}$ (with $s$ the
species, and $\nu_{s}$ the normalized viscosity) are considered at
this time. More accurate closures (such as gyroviscosity and anisotropic
heat transport) will be considered in future work.

Key to the character of the model in Eqs. \ref{eq:continuity-dimlss}-\ref{eq:temperature-dimlss}
is the specification of the electric field $\mathbf{E}$ via an Ohm's
law. In general, the Ohm's law is derived from the electron equation
of motion, which reads:
\begin{equation}
\mathbf{E}=-\mathbf{v}\times\mathbf{B}+\eta\mathbf{j}+\frac{d_{i}}{\rho}(\mathbf{j}\times\mathbf{B}-\nabla p_{e}-\nabla\cdot\Pi_{e})-\frac{d_{e}^{2}}{d_{i}}\frac{d\mathbf{v}_{e}}{dt}=-\mathbf{v}_{e}\times\mathbf{B}+\eta\mathbf{j}-\frac{d_{i}}{\rho}(\nabla p_{e}+\nabla\cdot\Pi_{e})-\frac{d_{e}^{2}}{d_{i}}\frac{d\mathbf{v}_{e}}{dt}.\label{eq:e_eom}
\end{equation}
The parameter $d_{i}$ characterizes the importance of Hall effects:
for $d_{i}\ll1$ the model transitions to resistive MHD, while for
$d_{i}\gg1$ the model transitions to the so-called electron MHD model
\citep{biskamp-book-00,emhd}. The parameter $d_{e}=c/\omega_{pe}L$
is the dimensionless electron inertial length, which characterizes
the importance of electron inertia. The ion and electron inertial
lengths are related as $d_{e}^{2}/d_{i}^{2}=m_{e}/m_{i}$.

The system of Eqs. \ref{eq:continuity-dimlss}-\ref{eq:temperature-dimlss},
together with Eq. \ref{eq:e_eom}, admits the following local energy
conservation law:
\[
\partial_{t}\left(\rho\frac{v^{2}}{2}+\frac{B^{2}}{2}+\frac{p}{\gamma-1}\right)+\nabla\cdot\left[\rho\frac{v^{2}}{2}\mathbf{v}+\mathbf{E}\times\mathbf{B}+\frac{\gamma p}{\gamma-1}\mathbf{v}^{*}+\sum_{s=i,e}\Pi_{s}\cdot\mathbf{v}_{s}+\mathbf{q}\right]=0.
\]
This implies that, with appropriate boundary conditions {[}e.g., perfect
conductor ($\mathbf{E}\times\mathbf{n}=0$), impenetrable wall ($\mathbf{v}\cdot\mathbf{n}=0$),
and no viscous stress ($\mathbf{n}\cdot\Pi_{s}=\mathbf{0}$){]}, the
total energy is exactly conserved. Numerically, energy will be conserved
only approximately, since we are solving for the specific internal
energy (Eq. \ref{eq:temperature-dimlss}) instead of total energy
(the former is more advantageous for low plasma-$\beta$ applications).
In our simulations, energy is typically conserved to better than one
part in $10^{3}$.

For sufficiently large $d_{i}$, the system in Eqs. \ref{eq:continuity-dimlss}-\ref{eq:temperature-dimlss},
together with Eq. \ref{eq:e_eom}, is dispersively hyperbolic, supporting
fast dispersive normal modes (Chap. 3 in Ref. \citep{swanson}). In
many applications of interest, however, dynamical time scales of interest
are much slower than those associated with these normal modes. This
is so because macroscopic dynamical time scales of interest are typically
related to the bulk plasma, and thus controlled by ions, while the
much faster normal modes are related to electron physics. In such
instances, an implicit approach that steps over the fast normal-mode
time scales to resolve the slower time scales of interest is advantageous.
This is the subject of this paper. However, as we shall see, the extreme
hyperbolic character of the Hall MHD model makes the task of developing
an optimal, scalable solver a very difficult task.

For preconditioning purposes, we will find it advantageous to consider
a governing equation for the magnetic vector potential ${\bf A}$
instead of the magnetic field $\mathbf{B}$ (which are related as
${\bf B}=\nabla\times{\bf A}$). Evolving $\mathbf{A}$ instead of
$\mathbf{B}$ has additional benefits in that it allows more flexibility
in the numerical discretization without sacrificing the solenoidal
property of the magnetic field. However, it is also more challenging
in that it requires a suitable gauge, and care must be taken with
boundary conditions. The evolution equation for ${\bf A}$ is found
from ${\bf E}=-\nabla\Phi-\partial_{t}{\bf A}$ (where $\Phi$ is
a scalar potential) and the generalized Ohm's law, Eq. \ref{eq:e_eom},
and reads: 
\begin{equation}
\frac{\partial{\bf A}}{\partial t}+\mathbf{E}=\frac{\partial{\bf A}}{\partial t}-\mathbf{v}_{e}\times\mathbf{B}+\eta\mathbf{j}-\frac{d_{i}}{\rho}(\nabla p_{e}+\nabla\cdot\Pi_{e})-\frac{d_{e}^{2}}{d_{i}}\frac{d\mathbf{v}_{e}}{dt}=-\nabla\Phi.
\end{equation}
One can eliminate $\Phi$ by choosing an appropriate gauge. Here,
we consider a variation of the Weyl gauge, $\Phi=-\frac{d_{e}^{2}}{d_{i}}\frac{\mathbf{v}_{e}^{2}}{2}$,
to find: 
\begin{equation}
\frac{\partial{\bf A}^{*}}{\partial t}-\mathbf{v}_{e}\times(\nabla\times{\bf A}^{*})+\eta\nabla\times\nabla\times{\bf A}-\frac{d_{i}}{\rho}(\nabla p_{e}+\nabla\cdot\Pi_{e})={\bf 0},\label{eqn: vec_pot_eqn}
\end{equation}
where $\mathbf{A}^{*}=\mathbf{A}-\frac{d_{e}^{2}}{d_{i}}\mathbf{v}_{e}$.
This equation is what we implement numerically. The solenoidal constraint
of the magnetic field is automatically guaranteed as long as $\nabla\cdot\nabla\times=0$
discretely (as is the case in our implementation \citep{chacon-cpc-04-mhd_discret}).
We discuss the specifics of the electron pressure tensor term in the
next section.

As stated earlier, a simple viscous closure is used for both species,
$\Pi_{s}\approx-\rho\nu_{s}\nabla\mathbf{v}_{s}$. It is important
to note that the main purpose of keeping the electron pressure tensor
contribution in our implementation, in the absence of a more informed
closure, is to regularize the electron equation of motion by providing
some dissipation scale. Such regularization is needed for numerical
purposes, as has been described in various other studies \citep{gem,chacon-JCP-hall}.
In our context, we have seen rapid degradation of the preconditioner
performance when such a term is lacking. The reason is that approximations
introduced in the preconditioner impact small scales the most \citep{chacon-pop-08-3dmhd},
and the dispersive nature of Hall MHD seems to amplify these differences
to the point that it renders the preconditioner ineffective after
just a few implicit time steps.

For electrons, the simple viscous closure gives:
\begin{equation}
\nabla\cdot\Pi_{e}=-\nabla\cdot(\rho\nu_{e}\nabla\mathbf{v}_{e})=-\nabla\cdot\left[\rho\nu_{e}\nabla\left(\mathbf{v}-\frac{d_{i}}{\rho}\mathbf{j}\right)\right]=-\nabla\cdot\left[\rho\nu_{e}\nabla\left(\mathbf{v}-\frac{d_{i}}{\rho}\nabla\times\nabla\times\mathbf{A}\right)\right],\label{eq:pi_e-closure-2}
\end{equation}
where $\nu_{e}$ is the electron viscosity (also known as hyperresistivity).
This expression contains a fourth-order dissipation operator on $\mathbf{A}$,
with $d_{i}\nabla\cdot\Pi_{e}\sim d_{i}^{2}\nu_{e}\nabla^{2}(\nabla\times\nabla\times\mathbf{A})$.
The higher differential order is needed to provide a dissipation length
scale to quadratic dispersive waves with dispersion relation $\omega\sim k^{2}$
\citep{gem}. Numerical considerations can help to determine a grid-bound
$\nu_{e}$ that provides sufficient dissipation while avoiding excessive
stiffness from the high-order operators. An estimate of $\nu_{e}$
can be obtained by balancing the damping rate $d_{i}^{2}\nu_{e}k^{4}$
with the dispersive wave frequency $d_{i}v_{A}k_{\parallel}k$, to
find \citep{chacon-JCP-hall}:
\begin{equation}
\nu_{e}=Cd_{i}^{-1}v_{A}k_{\parallel}k^{-3},\label{eq:hyperres-scaling}
\end{equation}
with $C$ a constant of order unity.

We comment briefly on the discretization of Eqs. \ref{eq:continuity-dimlss}-\ref{eq:temperature-dimlss}.
Spatially, the system is discretized using second-order finite volumes
with all variables (vectors and scalars) co-located at cell centers
\citep{chacon-cpc-04-mhd_discret}. Such spatial representation is
conservative and nonlinearly stable, and has proved to remarkably
robust both in resistive \citep{chacon-cpc-04-mhd_discret} and Hall
MHD (see Ref. \citep{kai-mrc-06} and results herein). Temporally,
we employ either a $\theta$-scheme with $\theta=0.5$ (second-order
Crank-Nicolson), or second-order backward differentiation formulas
\citep{byrne-acmtms-75-bdf}, depending on the problem. Either approach
results in a set of nonlinear algebraic equations $\mathbf{G}(\mathbf{x})=\mathbf{0}$,
with $\mathbf{x}^{T}=(\rho,T,\mathbf{A},\mathbf{v})$, that needs
to be inverted every time step. For this, we employ preconditioned
Jacobian-free Newton-Krylov (JFNK) methods. The next section discusses
our approach to preconditioning the Hall MHD system.

\section{Physics-based preconditioning strategy}

\label{sec:preconditioning}

We formulate our preconditioner by first parabolizing the implicitly
discretized hyperbolic system via an approximate block factorization,
and then invert the resulting systems with multigrid methods (MG)
for scalability.

As preconditioners, MG methods have been shown in many applications
of interest to lead to optimal convergence rates in JFNK \citep{knoll-jcp-04-nk_rev}.
The key element in a working MG solver is the smoother. While smoothers
can be found fairly easily for diagonally dominant systems (in a point
or block sense; p. 96 in \citep{wessMG}), it is remarkably hard otherwise.
Hyperbolic systems (such as MHD) can be shown to be diagonally submissive
when time steps larger than the explicit CFL stability constraint
are employed (see Ref. \citep{chacon-JCP-rmhd} for an in-depth explanation
of this issue). However, hyperbolic systems can be conveniently parabolized
in an implicit time-stepping setting. The basic idea is to produce
a well-conditioned (diagonally-dominant) parabolic operator from an
ill-conditioned hyperbolic system of equations by a block factorization
\citep{Elman-jcp-03-ns_schur,knoll-jcp-04-nk_rev,elman2006block,chacon-JCP-rmhd,chacon-JCP-hall,elman2008taxonomy,chacon-pop-08-3dmhd,cyr2012stabilization,cyr2013new,phillips2016block,bonilla2023fully}.
The procedure can be understood easily with a first-order coupled
hyperbolic linear system:
\begin{eqnarray*}
\partial_{t}u & = & \partial_{x}v,\\
\partial_{t}v & = & \partial_{x}u.
\end{eqnarray*}
Differencing implicitly in time (with backward Euler for simplicity)
but keeping the continuum spatial representation, we have:
\begin{eqnarray}
u^{n+1} & = & u^{n}+\Delta t\partial_{x}v^{n+1},\label{eq:hyp-1}\\
v^{n+1} & = & v^{n}+\Delta t\partial_{x}u^{n+1}.\label{eq:hyp-2}
\end{eqnarray}
It is now possible to substitute the second equation into the first
to obtain the following parabolic equation:
\begin{equation}
(\mathbb{I}-\Delta t^{2}\partial_{xx})u^{n+1}=u^{n}+\Delta t\partial_{x}v^{n},\label{eq:semi-implicit}
\end{equation}
which is equivalent to Eqs. \ref{eq:hyp-1}-\ref{eq:hyp-2}, but much
better conditioned because the parabolic operator is diagonally dominant.
Once $u^{n+1}$ is found, $v^{n+1}$ can be found straightforwardly
from Eq. \ref{eq:hyp-2}. A similar idea is behind the method of differential
approximations \citep{caramana-si}, and the semi-implicit solvers
developed in the MHD context \citep{harned-kerner-si,harned-schnack-si,harned-mikic-si}.

The parabolization of the hyperbolic equation is a manifestation of
the implicit discretization of the semi-discrete hyperbolic system.
Indeed, hyperbolic waves feature a finite propagation speed. By construction,
implicit methods are designed to step over stiff hyperbolic waves
to follow the dynamics on a slower time scale. In order to do so,
the implicit scheme must change the character of the equations for
unresolved hyperbolic time scales, and parabolization is the outcome.
Resolved hyperbolic time scales maintain their character. This can
be clearly seen from Eq. \ref{eq:semi-implicit}: for small enough
time steps, the second-order term can be neglected, and one recovers
an explicit (hyperbolic) formulation. It is only for large enough
time steps that the system changes character to become parabolic.

The connection between parabolization and the Schur factorization
\citep{chacon-JCP-hall,chacon-pop-08-3dmhd} enables the generalization
of these ideas to more complicated hyperbolic systems. Equations \ref{eq:hyp-1}-\ref{eq:hyp-2},
when written in block-matrix form, can be factorized as:
\[
\left[\begin{array}{cc}
\mathbb{I} & -\Delta t\partial_{x}\\
-\Delta t\partial_{x} & \mathbb{I}
\end{array}\right]=\left[\begin{array}{cc}
\mathbb{I} & -\Delta t\partial_{x}\\
0 & \mathbb{I}
\end{array}\right]\left[\begin{array}{cc}
\mathbb{I}-\Delta t^{2}\partial_{x}^{2} & 0\\
0 & \mathbb{I}
\end{array}\right]\left[\begin{array}{cc}
\mathbb{I} & 0\\
-\Delta t\partial_{x} & \mathbb{I}
\end{array}\right].
\]
The connection is now obvious, as the parabolic operator appears naturally.
Its usefulness will become apparent in the following sections.

\subsection{Block structure of the linearized Hall MHD model}

We begin by considering the linearized Hall MHD model for our preconditioning
development. Neglecting heat sources (which do not introduce stiff
time scales), the coupling structure of the linearized MHD system
in terms of the linear updates $\delta\rho$, $\delta T_{e}$, $\delta\mathbf{A}$,
and $\delta\mathbf{v}$ reads:
\begin{eqnarray*}
\delta\rho & = & \mathcal{L}_{\rho}(\delta\rho,\delta\mathbf{v}),\\
\delta p & = & \mathcal{L}_{T}(\delta p,\delta\mathbf{v},\delta\mathbf{A},\delta\rho),\\
\delta\mathbf{A} & = & \mathcal{L}_{E}(\delta\mathbf{v},\delta\mathbf{A},\delta\rho,\delta p),\\
\delta\mathbf{v} & = & \mathcal{L}_{v}(\delta\mathbf{v},\delta\mathbf{A},\delta\rho,\delta p),
\end{eqnarray*}
where the $\mathcal{L}_{i}$ represent linear operators. The resulting
Jacobian structure is:

\begin{equation}
J\delta\mathbf{x}=\left[\begin{array}{cccc}
D_{\rho} & 0 & 0 & U_{\rho\mathbf{v}}\\
\myBox{L_{p\rho}} & D_{p} & \myBox{U_{p\mathbf{A}}} & U_{p\mathbf{v}}\\
\myBox{L_{\mathbf{A}\rho}} & \myBox{L_{\mathbf{A}p}} & \myBox{D_{\mathbf{A}}} & U_{\mathbf{vA}}\\
L_{\mathbf{v}\rho} & L_{\mathbf{v}p} & L_{\mathbf{v}\mathbf{A}} & D_{\mathbf{v}}
\end{array}\right]\left(\begin{array}{c}
\delta\rho\\
\delta p\\
\delta\mathbf{A}\\
\delta\mathbf{v}
\end{array}\right).\label{eq:Jacobian}
\end{equation}
The boxed blocks contain Hall-MHD contributions from finite $d_{i}$
terms (stemming from the Ohm's law and the definition of $\mathbf{v}^{*}$
in the temperature equation). Here, diagonal blocks are given by:
\begin{eqnarray*}
D_{\rho}\delta\rho & = & \frac{\delta\rho}{\Delta t}+\theta[\nabla\cdot(\mathbf{v}_{0}\delta\rho)-D\nabla^{2}\delta\rho],\\
D_{p}\delta p & = & \left(\frac{1}{\Delta t}+\theta(\gamma-1)\nabla\cdot\mathbf{v}_{0}^{*}\right)\delta p+\theta[\nabla\cdot(\mathbf{v}_{0}^{*}\delta p)-\frac{\kappa}{\alpha_{T}\rho_{0}}\nabla^{2}\delta T],\\
D_{\mathbf{A}}\delta\mathbf{A} & = & \frac{\delta\mathbf{A}^{*}}{\Delta t}-\theta\left(\mathbf{v}_{e,0}\times\nabla\times\delta\mathbf{A}^{*}-d_{i}\frac{\nabla\times\nabla\times\delta\mathbf{A}}{\rho_{0}}\times\mathbf{B_{0}}-\eta\nabla\times\nabla\times\delta\mathbf{A}+\nabla\cdot\left[\rho_{0}d_{i}\nu_{e}\nabla\left(\frac{\nabla\times\nabla\times\delta\mathbf{A}}{\rho_{0}}\right)\right]\right),\\
D_{\mathbf{v}}\delta\mathbf{v} & = & \rho_{0}\left[\frac{\delta\mathbf{v}}{\Delta t}+\theta(\mathbf{v}_{0}\cdot\nabla\delta\mathbf{v}+\delta\mathbf{v}\cdot\nabla\mathbf{v}_{0})\right]-\theta\nabla\cdot(\nu_{i}\rho_{0}\nabla\delta\mathbf{v}),
\end{eqnarray*}
where, for $D_{\mathbf{v}}$, we have used the non-conservative form
of the momentum equation \citep{chacon-pop-08-3dmhd}. Here, $\theta$
is the time centering parameter, and $\delta\mathbf{A}^{*}=\delta\mathbf{A}+\frac{d_{e}^{2}}{\rho_{0}}\nabla\times\nabla\times\delta\mathbf{A}$.
Off-diagonal blocks $L$ and $U$ are given by:
\begin{eqnarray*}
U_{\rho\mathbf{v}}\delta\mathbf{v} & = & \theta\nabla\cdot(\delta\mathbf{v}\rho_{0}),\\
U_{p\mathbf{A}}\delta\mathbf{A} & = & -\theta d_{i}\left[\nabla\cdot(T_{e,0}\nabla\times\nabla\times\delta\mathbf{A})+(\gamma-1)p_{e,0}\nabla\cdot\frac{\nabla\times\nabla\times\delta\mathbf{A}}{\rho_{0}}\right],\\
U_{\mathbf{vA}}\delta\mathbf{v} & = & -\theta\delta\mathbf{v}\times\mathbf{B}_{0},\\
U_{p\mathbf{v}}\delta\mathbf{v} & = & \theta[\nabla\cdot(\delta\mathbf{v}p_{0})+(\gamma-1)p_{0}\nabla\cdot\delta\mathbf{v}],\\
L_{\mathbf{A}\rho}\delta\rho & \approx & \theta d_{i}\frac{\delta\rho}{\rho_{0}^{2}}\nabla p_{e,0},\\
L_{p\rho}\delta\rho & = & \theta\frac{d_{i}}{1+\alpha}\left[\nabla\cdot\left(\frac{\delta\rho}{\rho_{0}^{2}}p_{0}\mathbf{j}_{0}\right)+(\gamma-1)p_{0}\nabla\cdot\left(\frac{\delta\rho}{\rho_{0}^{2}}\mathbf{j}_{0}\right)\right],\\
L_{\mathbf{A}p}\delta p & = & -\theta\frac{d_{i}}{1+\alpha}\frac{1}{\rho_{0}}\nabla\delta p,\\
L_{\mathbf{v}\rho}\delta\rho & = & \delta\rho\left(\frac{\mathbf{v}_{0}}{\Delta t}+\theta\mathbf{v}_{0}\cdot\nabla\mathbf{v}_{0}\right)-\theta\nabla\cdot((\nu_{i}+\nu_{e})\delta\rho\nabla\mathbf{v}_{0}),\\
L_{\mathbf{vA}}\delta\mathbf{A} & = & -\theta\left[\mathbf{j}_{0}\times\nabla\times\delta\mathbf{A}+(\nabla\times\nabla\times\delta\mathbf{A})\times\mathbf{B}_{0}\right],\\
L_{\mathbf{v}p}\delta p & = & \theta\nabla\delta p.
\end{eqnarray*}

\subsection{Parabolization of the Hall MHD Jacobian matrix}

Upon inspection of Eq. \ref{eq:Jacobian}, it is clear that the the
convenient ``arrow structure'' of the resistive MHD Jacobian matrix
\citep{chacon-pop-08-3dmhd} has been spoiled. However, the stiff
electron timescales introduced in Hall MHD are all contained in the
magnetic field diagonal block $D_{\mathbf{A}}$ via the $d_{i}(\frac{\nabla\times\nabla\times\delta\mathbf{A}}{\rho_{0}}\times\mathbf{B_{0}})$
term, responsible for the propagation of fast whistler waves. This
realization opens the way for a tractable physics-based preconditioning
algorithm.

The Jacobian recovers a manageable structure by simply neglecting
the block $U_{p\mathbf{A}}$ (stemming from the linearization of the
current correction in the $\mathbf{v}_{s}$ velocity, which is not
stiff). Then, the approximate Jacobian system reads:
\[
J\delta\mathbf{x}\approx\left[\begin{array}{cc}
M & U\\
L & D_{\mathbf{v}}
\end{array}\right]\left(\begin{array}{c}
\delta\mathbf{y}\\
\delta\mathbf{v}
\end{array}\right),
\]
where $\delta\mathbf{y}=(\delta\rho,\delta\mathbf{A},\delta p)^{T}$,
and 
\begin{equation}
M\approx\left(\begin{array}{ccc}
D_{\rho} & 0 & 0\\
L_{p\rho} & D_{p} & 0\\
L_{\mathbf{A}\rho} & L_{\mathbf{A}p} & D_{\mathbf{A}}
\end{array}\right)\label{eq:approx_M}
\end{equation}
a lower triangular system. The exact Schur factorization of the inverse
of the 2$\times$2 block Jacobian matrix yields:

\[
\left[\begin{array}{cc}
M & U\\
L & D_{\mathbf{v}}
\end{array}\right]^{-1}=\left[\begin{array}{cc}
\mathbb{I} & -M^{-1}U\\
0 & \mathbb{I}
\end{array}\right]\left[\begin{array}{cc}
M^{-1} & 0\\
0 & P_{Schur}^{-1}
\end{array}\right]\left[\begin{array}{cc}
\mathbb{I} & 0\\
-LM^{-1} & \mathbb{I}
\end{array}\right],
\]
where $P_{Schur}=D_{\mathbf{v}}-LM^{-1}U$ is the Schur complement,
which contains all the information from the off-diagonal blocks $L$
and $U$ (similarly to the simple coupled-wave-equation example discussed
earlier). This gives the \emph{exact} inversion process:
\begin{eqnarray}
\delta\mathbf{y}^{*} & = & -M^{-1}\mathbf{G_{y}},\nonumber \\
\delta\mathbf{v} & = & P_{Schur}^{-1}[-\mathbf{G_{v}}-LM^{-1}\delta\mathbf{y}^{*}],\label{eq:exact_Schur}\\
\delta\mathbf{y} & = & \delta\mathbf{y}^{*}-\Delta t\,M^{-1}U\delta\mathbf{v}.\nonumber 
\end{eqnarray}

As it stands, however, inverting $P_{Schur}$ is impractical due to
the presence of $M^{-1}$. Suitable simplifications have been proposed
for small ion flows \citep{chacon-pop-08-3dmhd}, and later generalized
to arbitrary ion flows \citep{chacon-jpcs-08-scidac}, which cannot
yet be directly applied in this context owing to the presence of the
very stiff vector-potential diagonal block, $D_{\mathbf{A}}$. However,
progress can be made by realizing that the vector potential diagonal
block is the only one supporting stiff EMHD physics, while the rest
of the algorithm in Eq. \ref{eq:exact_Schur} is correcting for ion
dynamics. Since electron Hall physics is largely decoupled from ions,
we hypothesize that an operator-split approach in which the electrons
response is solved for first, followed by an ion response correction
will form the basis for an effective preconditioner.

Making this approximation explicit in Eq. \ref{eq:exact_Schur}, we
perform a full solve of $M^{-1}$ in the first step, while in subsequent
ones we neglect electron dynamics and approximate $M^{-1}\approx\Delta t\mathbb{I}$
(small-bulk-flow approximation \citep{chacon-pop-08-3dmhd}). The
resulting approximate ``physics-based'' preconditioner reads:
\begin{eqnarray}
\delta\mathbf{y}^{*} & = & -M^{-1}\mathbf{G_{y}},\nonumber \\
\delta\mathbf{v} & = & P_{SF}^{-1}[-\mathbf{G_{v}}-L\delta\mathbf{y}^{*}],\label{eq:approx_Schur}\\
\delta\mathbf{y} & = & \delta\mathbf{y}^{*}-\Delta t\,U\delta\mathbf{v},\nonumber 
\end{eqnarray}
with $P_{SF}=D_{\mathbf{v}}-\Delta t\,LU$. This is formally identical
to the resistive MHD preconditioner proposed in \citep{chacon-pop-08-3dmhd},
except now the inversion of the $M$ block contains the EMHD block,
which is the stiffest in the system and whose treatment we discuss
in detail below. As in the reference, taking $L$ and $U$ from the
linearized form of Eqs. \ref{eq:continuity-dimlss}-\ref{eq:temperature-dimlss},
using the non-conservative form of Eq. \ref{eq:EOM-dimlss}, and Picard-linearizing
the density (by ignoring $\delta\rho$ terms), the operator $P_{SF}$
acting on $\delta\mathbf{v}$ reads:

\begin{equation}
P_{SF}\delta\mathbf{v}=\left[\rho_{0}\frac{\delta\mathbf{v}}{\Delta t}+\theta\rho_{0}(\mathbf{v}_{0}\cdot\nabla\delta\mathbf{v}+\delta\mathbf{v}\cdot\nabla\mathbf{v}_{0})-\theta\nabla\cdot(\rho_{0}\nu_{i}\nabla\delta\mathbf{v})\right]+\Delta t\theta^{2}W(\mathbf{B}_{0},p_{0})\delta\mathbf{v},\label{eq:schur-comp}
\end{equation}
where $W(\mathbf{B}_{0},p_{0})$ results from the composition of off-diagonal
operators $L_{\mathbf{A}\mathbf{v}}$ and $U_{\mathbf{vA}}$, and
reads: 
\begin{equation}
W(\mathbf{B}_{0},p_{0})\delta\mathbf{v}\approx\mathbf{B}_{0}\times\nabla\times\nabla\times[\delta\mathbf{v}\times\mathbf{B}_{0}]-\mathbf{j}_{0}\times\nabla\times[\delta\mathbf{v}\times\mathbf{B}_{0}]-\nabla[\delta\mathbf{v}\cdot\nabla p_{0}+\gamma p_{0}\nabla\cdot\delta\mathbf{v}],\label{eq:energy_op}
\end{equation}
with $\mathbf{j}_{0}=\nabla\times\mathbf{B}_{0}$. This is a linearized
form of the MHD energy principle, which is self-adjoint when $\mathbf{j}_{0}\times\mathbf{B}_{0}=\nabla p_{0}$
(and therefore has real eigenvalues in that case). Thus, the MHD system
has been effectively parabolized.

Equations \ref{eq:approx_Schur}, with $M$ defined in Eq. \ref{eq:approx_M},
is the proposed preconditioner. It reuses most of the elements of
the resistive MHD preconditioner proposed in Ref. \citep{chacon-pop-08-3dmhd},
except for the treatment of the vector potential diagonal block, which
we discuss next.

\section{Numerical implementation}

\label{sec:Numerical-details}

We employ a Newton-FGMRES (flexible GMRES \citep{saad-fgmres}) nonlinear
iterative method, implemented Jacobian-free. Newton nonlinear convergence
is controlled in our implementation by the usual criterion:
\begin{equation}
\left\Vert \mathbf{G}(\mathbf{x}_{k})\right\Vert _{2}<\epsilon_{a}+\epsilon_{r}\left\Vert \mathbf{G}(\mathbf{x}_{0})\right\Vert _{2}=\epsilon_{t},\label{eq-Newton-conv-tol-1}
\end{equation}
where $\left\Vert \cdot\right\Vert _{2}$ is the $L_{2}$-norm (euclidean
norm), $\epsilon_{a}=\sqrt{N}\times10^{-15}$ (with $N$ the total
number of degrees of freedom) is an absolute tolerance to avoid converging
below round-off, $\epsilon_{r}$ is the Newton relative convergence
tolerance (set to $10^{-3}$ in this work), and $\mathbf{G}(\mathbf{x}_{0})$
is the initial residual. GMRES convergence is controlled by an inexact
Newton method \citep{inexact-newton} in which the Krylov convergence
tolerance every Newton iteration is adjusted as follows:
\begin{equation}
\left\Vert J_{k}\delta\mathbf{x}_{k}+\mathbf{G}(\mathbf{x}_{k})\right\Vert _{2}<\zeta_{k}\left\Vert \mathbf{G}(\mathbf{x}_{k})\right\Vert _{2}\label{eq-inexact-newton-1}
\end{equation}
where $\zeta_{k}$ is the inexact Newton parameter and $J_{k}=\left.\frac{\partial\mathbf{G}}{\partial\mathbf{x}}\right|_{k}$
is the Jacobian matrix. Here, we employ the same prescription for
$\zeta_{k}$ as in earlier studies \citep{chacon-JCP-hall,chacon-pop-08-3dmhd}:
\begin{eqnarray*}
\zeta_{k}^{A} & = & \gamma\left(\frac{\left\Vert \mathbf{G}(\mathbf{x}_{k})\right\Vert _{2}}{\left\Vert \mathbf{G}(\mathbf{x}_{k-1})\right\Vert _{2}}\right)^{\alpha},\\
\zeta_{k}^{B} & = & \min[\zeta_{max},\max(\zeta_{k}^{A},\gamma\zeta_{k-1}^{\alpha})],\\
\zeta_{k} & = & \min[\zeta_{max},\max(\zeta_{k}^{B},\gamma\frac{\epsilon_{t}}{\left\Vert \mathbf{G}(\mathbf{x}_{k})\right\Vert _{2}})],
\end{eqnarray*}
with $\alpha=1.5$ , $\gamma=0.9$, and $\zeta_{max}=0.8$. The convergence
tolerance $\epsilon_{t}$ is the same as in Eq. \ref{eq-Newton-conv-tol-1}.
In this prescription, the first step ensures superlinear convergence
(for $\alpha>1$), the second avoids volatile decreases in $\zeta_{k}$,
and the last avoids oversolving in the last Newton iteration.

Our geometric MG solver is fully parallel, with coarse-grid parallel
domain agglomeration \citep{hulsemann2006parallel,petsc-mg-parallel-2016}.
It features a matrix-light implementation \citep{chacon-JCP-hall,chacon-pop-08-3dmhd},
in which only the diagonal of the system of interest is stored for
smoothing purposes. Coarse operators are found via rediscretization,
and required residuals in the MG iteration are found in a matrix-free
manner. As a smoother, we employ three passes of damped Jacobi (p.
10 in \citep{briggs-MG}; p. 118 in \citep{wessMG}), with weight
$\omega=0.7$, for both the restriction and the prolongation steps.
MG restriction employs conservative agglomeration, and prolongation
employs a first-order interpolation. In MG jargon, such V-cycle is
identified as V(3,3), where the two integers indicate restriction
and prolongation smoothing steps, respectively. In parallel, we use
colored damped Jacobi, which parallelizes very well; four colors are
needed in 2D (for a 9-point stencil), and 8 colors in 3D (for a 27-point
stencil). Our matrix-light geometric-MG approach features excellent
parallel performance, with wall-clock time scaling logarithmically
with the number of MPI tasks (as demonstrated in the next section).

For typical timesteps in HMHD, a single V(3,3) MG V-cycle is sufficient
to approximately invert the diagonal blocks $D_{\rho}$, $D_{p}$,
and $P_{SF}$ in Eq. \ref{eq:approx_Schur}. However, the EMHD block
$D_{\mathbf{A}}$ is much more challenging to invert robustly. For
this block, we use MG-preconditioned GMRES, converged to either 10
Krylov iterations or a relative tolerance of $0.1$, whichever comes
first (note that nested GMRES solves in the preconditioner are allowed
by FGMRES). This strategy still requires a working MG solver for the
EMHD block, which in turn requires a suitable smoother. We describe
our smoothing strategy for this block next.

\subsection{Smoothing of the vector-potential diagonal block}

In order to make the $D_{\mathbf{A}}$ operator MG-friendly we need
to provide a suitable smoother. Here, we are seeking to use damped
block-Jacobi smoothing, which is cheap and easily parallelizable.
The $D_{\mathbf{A}}$ diagonal block contains fast electron hyperbolic
physics (whistler waves, with dispersion relation $\omega\sim d_{i}v_{A}kk_{\parallel}$),
electron inertia, and electron viscosity (also known as hyperresistivity),
and reads:
\begin{eqnarray*}
D_{\mathbf{A}}\delta\mathbf{A} & = & \frac{\delta\mathbf{A}^{*}}{\Delta t}\\
 & - & \theta\left(\mathbf{v}_{e,0}\times\nabla\times\delta\mathbf{A}^{*}-d_{i}\underbrace{\frac{\nabla\times\nabla\times\delta\mathbf{A}}{\rho_{0}}\times\mathbf{B_{0}}}_{\mathrm{Whistler\,term}}-\eta\nabla\times\nabla\times\delta\mathbf{A}+\underbrace{\nabla\cdot\left[\rho_{0}d_{i}^{2}\nu_{e}\nabla\left(\frac{\nabla\times\nabla\times\delta\mathbf{A}}{\rho_{0}}\right)\right]}_{\mathrm{Hyperresistivity\,term}}\right),
\end{eqnarray*}
with:
\[
\delta\mathbf{A}^{*}=\delta\mathbf{A}+\frac{d_{e}^{2}}{\rho_{0}}\nabla\times\nabla\times\delta\mathbf{A}.
\]
This operator is the linearized form of the EMHD model.

The iterative solution of the vector-potential diagonal block is extremely
challenging, and has in fact been the key roadblock overcome in this
research. It is a stiff, high differential order, vectorial system
of equations. It supports fast (quadratic) dispersive waves, and features
a fourth-order dissipative operator of the form $\nabla^{2}(\nabla\times\nabla\times)$
that provides a dissipation length scale to the fast dispersive waves,
needed for nonlinear stability of the discretization \citep{gem}.
The form of the dissipative operator is not only problematic because
of its high order, but also because it features an infinitely degenerate
null space: any gradient component in the vector potential will get
annihilated by the curl. Although the $D_{\mathbf{A}}$ operator in
aggregate is non-singular, the presence of a null space in the hyperresistivity
term makes its MG implementation problematic. One may consider approximating
it, but being the term with the highest differential order, our experience
is that any approximations introduced to it will negatively (and significantly)
impact the performance of the preconditioner. In particular, replacing
$\nabla\times\nabla\times$ by $-\nabla^{2}$ is not justified in
principle because we do not use the Coulomb gauge ($\nabla\cdot\mathbf{A}=0$)
but a modified Weyl gauge, as discussed earlier. (A Coulomb gauge
could in principle be used, but it would result in an additional equation
for the electrostatic potential, which the Weyl gauge avoids.)

To provide for a suitable smoother, we begin by reformulating the
EMHD system by splitting it into two second-order vectorial systems:
\begin{eqnarray}
\frac{1}{\Delta t}\left(\delta\mathbf{A}+d_{e}^{2}\frac{\delta\mathbf{j}}{\rho_{0}}\right)-\theta\left(\mathbf{v}_{e,0}\times\nabla\times\left(\delta\mathbf{A}+d_{e}^{2}\frac{\delta\mathbf{j}}{\rho_{0}}\right)-d_{i}\frac{\delta\mathbf{j}}{\rho_{0}}\times\mathbf{B_{0}}-\eta\delta\mathbf{j}+\nabla\cdot\left[\rho_{0}d_{i}^{2}\nu_{e}\nabla\left(\frac{\delta\mathbf{j}}{\rho_{0}}\right)\right]\right) & = & \mathbf{rhs},\nonumber \\
\delta\mathbf{j}-\nabla\times\nabla\times\delta\mathbf{A} & = & \mathbf{0}.\label{eq:aux_dj}
\end{eqnarray}
The auxiliary variable $\delta\mathbf{j}$ is a physical quantity
(current density), with a well-defined set of boundary conditions,
ensuring the split system is well posed. The feasibility of this physical
system splitting without approximation is, in fact, the key reason
why we pursued a vector-potential-based representation of our Hall
MHD model instead of a magnetic-field-based one. For the latter, the
hyperresistivity term would read:
\[
\nabla\times\nabla\cdot\left[\rho_{0}d_{i}^{2}\nu_{e}\nabla\left(\frac{\nabla\times\delta\mathbf{B}}{\rho_{0}}\right)\right],
\]
which \emph{cannot} be split into two second-order systems in an obvious
way.

The Jacobi smoother for the coupled system in Eq. \ref{eq:aux_dj}
iterates on all local vector components for $\delta\mathbf{A}$ and
$\delta\mathbf{j}$ simultaneously on a per-cell basis. To deal with
the $\nabla\times\nabla\times$ operator in the smoothing stage, we
employ a defect-correction approach whereby we selectively approximate
$\nabla\times\nabla\times$ by $-\nabla^{2}$ in the computation of
the diagonal blocks, but retain the $\nabla\times\nabla\times$ form
for all the residual computations within the MG V-cycle. We show in
the Appendix for the special case of a homogeneous plasma that this
strategy produces a convergent damped-Jacobi iteration with the smoothing
property (critical for the effectiveness of MG), and therefore provides
an excellent basis for an effective MG preconditioner.

\section{Numerical results}

\label{sec:Numerical-results}

We have implemented this algorithm in the PIXIE3D nonlinear MHD code
\citep{chacon-cpc-04-mhd_discret,chacon-pop-08-3dmhd,chacon-jpcs-08-scidac}.
In what follows, we present numerical examples that demonstrate the
correctness of our implementation as well as its performance (algorithmically
and in parallel). We consider both linear examples (wave propagation)
and nonlinear examples (with the well-known GEM challenge magnetic
reconnection problem \citep{gem}, which we verify against the HiFi
nonlinear MHD code~\citep{lukin2016overview}). The latter example
is also considered for the algorithmic and parallel performance assessment.
The GEM challenge problem benefits from an implicit treatment, since
the dynamical time scale of interest is independent of the grid resolution,
and much slower than fast HMHD normal modes.

\subsection{Linear verification: whistler wave}

\begin{figure}
\begin{centering}
\includegraphics[width=0.7\columnwidth]{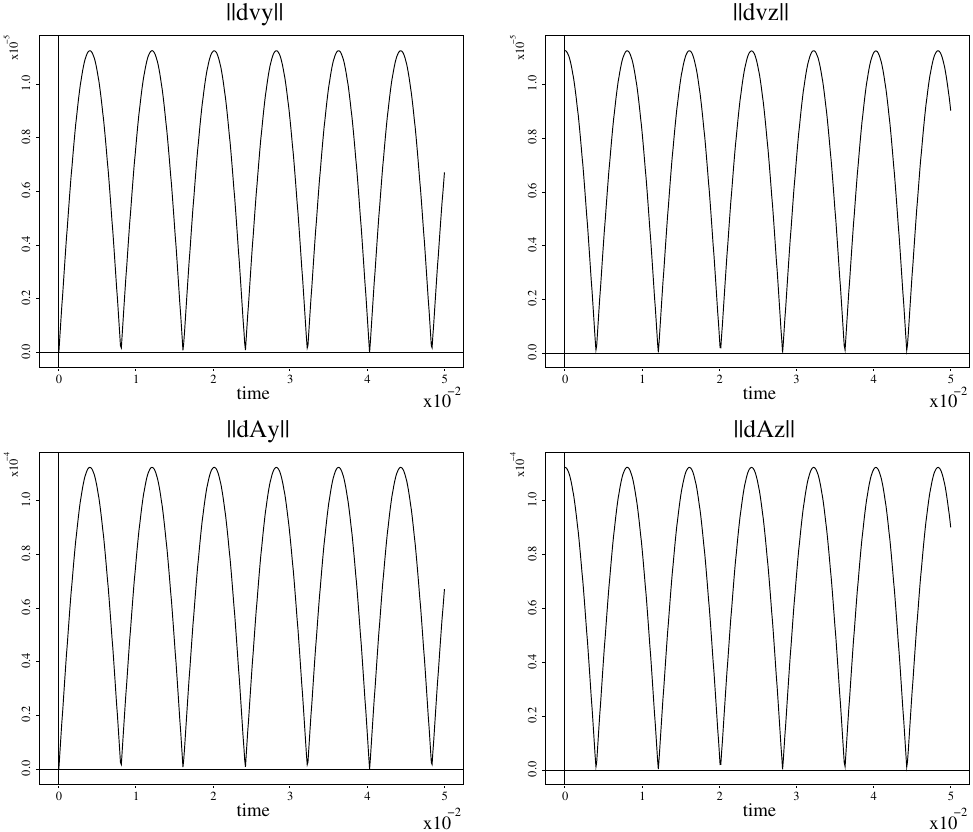}
\par\end{centering}
\caption{\label{fig:whistler-verification}Time histories of perturbations
of $v_{y}$, $v_{z}$, $A_{y}$, $A_{z}$ for the whistler wave linear
verification test.}
\end{figure}

We consider a collisionless, homogeneous, adiabatic ($\gamma=1$)
plasma with $\rho_{0}=1$, $T_{i0}=T_{e0}=1$, $\mathbf{B}_{0}=(1,0,0)$,
$\mathbf{v}=\mathbf{0}$, and $d_{i}=10$ in a one-dimensional domain
with $L_{x}=1$. Linear theory predicts that the frequency of the
resulting wave is $\omega=\pm d_{i}k_{x}^{2}$, with mode structure:
\begin{eqnarray*}
\delta A_{y} & = & -i\delta A_{z},\\
\omega\delta v_{y} & = & k_{x}^{2}\delta A_{y},\\
\omega\delta v_{z} & = & k_{x}^{2}\delta A_{z}.
\end{eqnarray*}
For $k_{x}=2\pi/L_{x}=2\pi$, the wave period is $T=2\pi/\omega=(2\pi d_{i})^{-1}\approx0.016.$
We set up a standing whistler wave with a perturbation of the form:
\begin{eqnarray*}
\delta A_{z} & = & -\frac{\epsilon}{k_{x}}\sin(k_{x}x),\\
\delta v_{z} & = & \frac{\epsilon}{k_{x}d_{i}}\sin(k_{x}x),
\end{eqnarray*}
with $\epsilon=10^{-3}$. Simulations are performed using 32 mesh
points and $\Delta t=10^{-4}$. Figure \ref{fig:whistler-verification}
shows the time histories of the $\ell_{2}$-norm of $\delta v_{y}$,
$\delta v_{z}$, $\delta A_{y}$, $\delta A_{z}$, where one can appreciate
that, as expected: 1) wave period is 0.016, 2) $\left\Vert \delta v_{y}\right\Vert $
and $\left\Vert \delta A_{y}\right\Vert $ are in phase, as are $\left\Vert \delta v_{z}\right\Vert $
and $\left\Vert \delta A_{z}\right\Vert $, and 3) $\left\Vert \delta A_{y}\right\Vert $
and $\left\Vert \delta A_{z}\right\Vert $ are 90 degrees out of phase
(as are $\left\Vert \delta v_{y}\right\Vert $ and $\left\Vert \delta v_{z}\right\Vert $).

\subsection{Linear verification: kinetic Alfvén wave}

We consider a collisionless, homogeneous, adiabatic ($\gamma=1$)
plasma with $\rho_{0}=1$, $T_{i0}=T_{e0}=2.5\times10^{3}$, $\mathbf{B}_{0}=(1,0,0)$,
$\mathbf{v}=\mathbf{0}$, and $d_{i}=5/\pi$ in a two-dimensional
domain with $L_{x}=10$ and $L_{y}=1$. Linear theory predicts that
the frequency of the resulting wave is $\omega\approx\pm\rho_{s}k_{x}k_{y}$,
with $\rho_{s}=d_{i}\sqrt{\beta/2}$ the sound Larmor radius and $\beta=2\mu_{0}p_{0}/B_{0}^{2}=10^{4}$
the plasma beta, and mode structure:
\begin{eqnarray}
\delta A_{z} & = & \frac{i}{\rho_{s}k_{x}^{2}}\delta v_{y},\label{eq:kaw-mode-a}\\
\delta v_{y} & = & \frac{k_{x}}{k_{y}}[d_{i}^{2}(k_{x}^{2}+k_{y}^{2})-1]\delta v_{x}.\label{eq:kaw-mode-v}
\end{eqnarray}
For $k_{x}=2\pi/L_{x}=\pi/5$, $k_{y}=2\pi n_{y}/L_{y}=2\pi n_{y}$,
and $n_{y}=5$, the wave period is $T=2\pi/\omega=n_{y}^{-1}\sqrt{2/\beta}\approx2.83\times10^{-3}.$
We set up a standing wave with a perturbation of the form:
\begin{eqnarray*}
\delta v_{x} & = & \epsilon\cos(k_{x}x+k_{y}y),\\
\delta v_{y} & = & \epsilon\frac{k_{x}}{k_{y}}[d_{i}^{2}(k_{x}^{2}+k_{y}^{2})-1]\cos(k_{x}x+k_{y}y),
\end{eqnarray*}
with $\epsilon=10^{-7}$. Simulations are performed using 32$\times$128
mesh points and $\Delta t=4\times10^{-5}$. Figure \ref{fig:whistler-verification}
shows the time histories of the $\ell_{2}$-norm of $\delta v_{y}$,
$\delta v_{z}$, $\delta A_{y}$, $\delta A_{z}$, where one can appreciate
that, as expected: 1) wave period is $\approx2.8\times10^{-3}$, 2)
$\left\Vert \delta v_{x}\right\Vert $ and $\left\Vert \delta v_{y}\right\Vert $
are in phase and the ratio in magnitude is $\delta v_{y}/\delta v_{x}\approx50$
(as expected from Eq. \ref{eq:kaw-mode-v}), and 3) $\left\Vert \delta A_{z}\right\Vert $
is 90 degrees out of phase from the velocity perturbations, and consistent
in magnitude with Eq. \ref{eq:kaw-mode-a}.
\begin{figure}
\begin{centering}
\includegraphics[width=1\columnwidth]{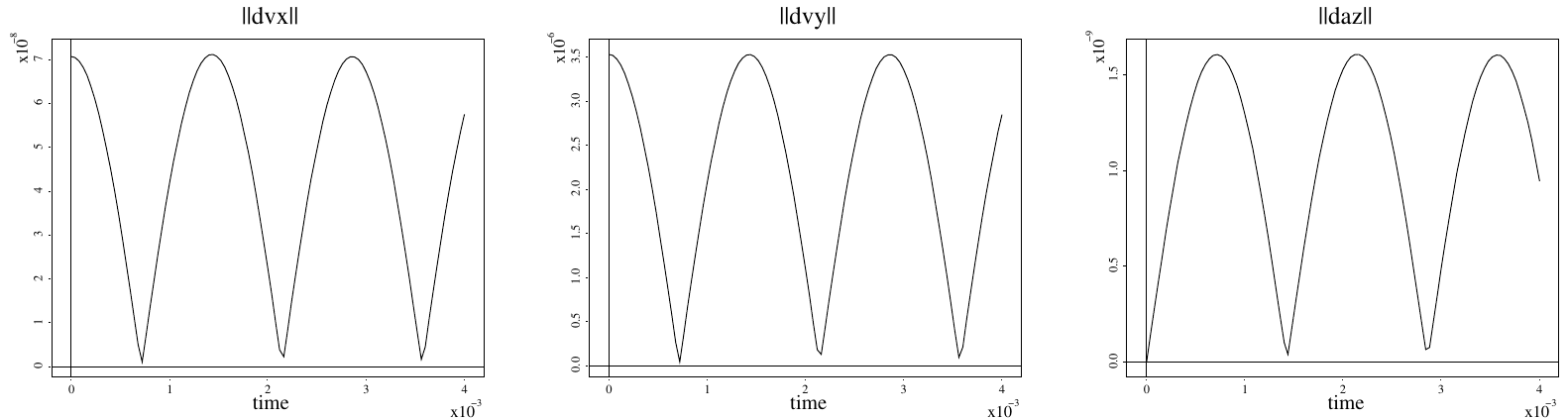}
\par\end{centering}
\caption{\label{fig:whistler-verification-1}Time histories of perturbations
of $v_{x}$, $v_{y}$, $A_{z}$ for the KAW wave linear verification
test.}
\end{figure}

\subsection{Nonlinear verification: GEM challenge magnetic reconnection problem}

\begin{figure}
\begin{centering}
\includegraphics[width=1\columnwidth]{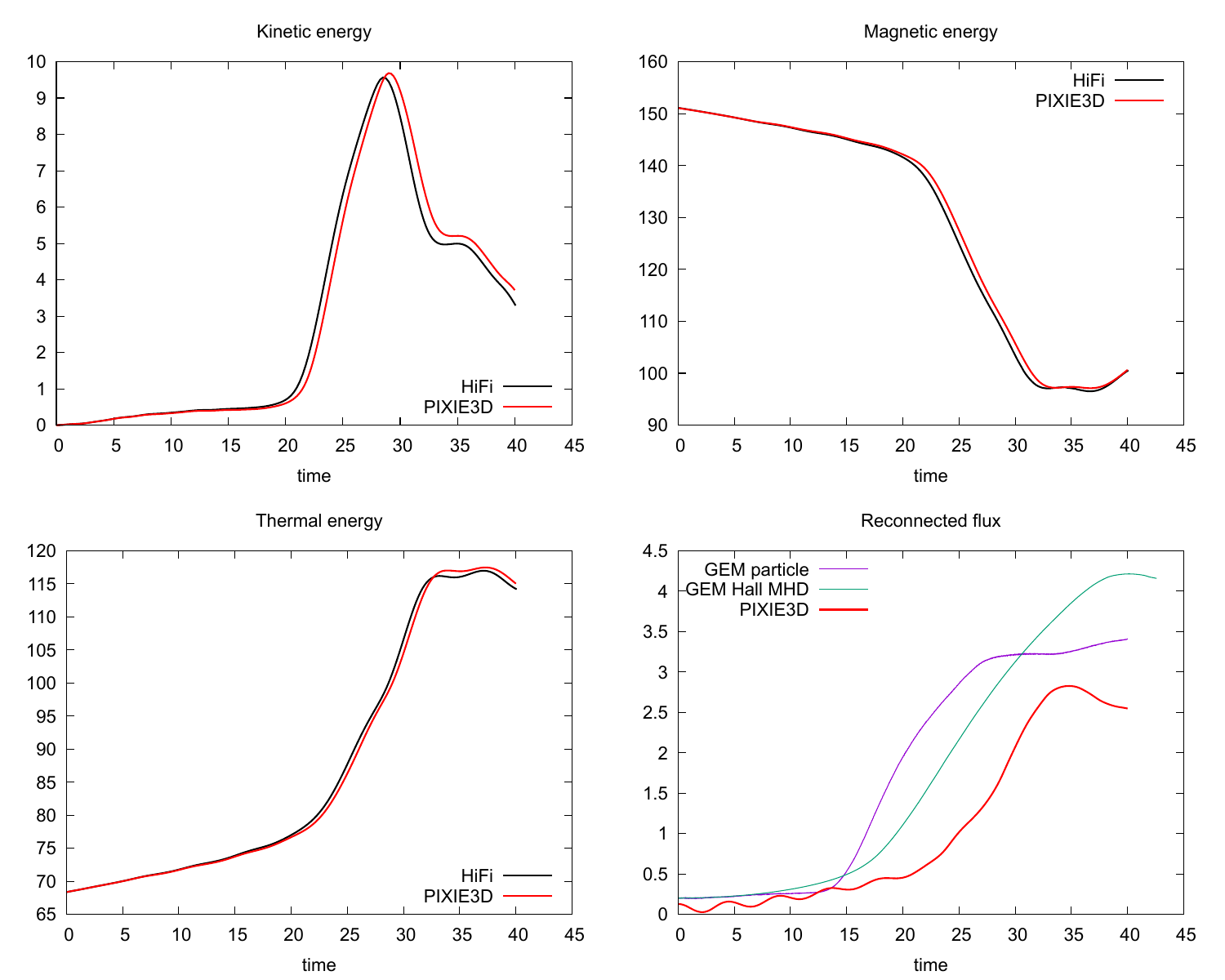}
\par\end{centering}
\caption{\label{fig:hifi_vs_pixie3d_gem}Nonlinear comparison of the temporal
evolution of the kinetic, magnetic and thermal energies from the PIXIE3D
and HiFi codes for the GEM challenge (see text for setup). The reconnected
flux for PIXIE3D is compared with data from Fig. 1 of Ref. \citep{gem},
with excellent agreement in the slope of the curve (which determines
the peak reconnection rate) vs. the GEM results.}
\end{figure}
 We consider the GEM challenge problem \citep{gem} for a nonlinear
verification study against the HiFi nonlinear extended MHD code. HiFi
\citep{lukin2016overview} is a highly accurate, fully implicit, spectral-element-based
nonlinear extended MHD code. To preserve smoothness during the simulation,
HiFi features a moving mesh strategy based on a Laplace-Beltrami mesh
governing equation (see Ref. \citep{chacon-jcp-05-adgrid-hf} and
references therein). For consistency with HiFi, we have replaced $\mathbf{v}^{*}$
in Eq. \ref{eq:temperature-dimlss} with $\mathbf{v}$.

The GEM challenge problem setup is a Harris current sheet, defined
by the vector potential $(0,0,A_{z})$, with:
\[
A_{z}=-\lambda\ln(\cosh(x/\lambda)),
\]
with $\lambda=0.5\,d_{i}$, balanced by a pressure profile given by
constant ion and electron temperatures with $T_{i}/T_{e}=\alpha=5$
and $(T_{i}+T_{e})=1/2$, and density profile:
\[
\rho(x)=\frac{n_{0}}{\cosh(x/\lambda)}+n_{\infty}.
\]
with $n_{0}=1$, $n_{\infty}=0.2$. We consider a slab domain $(-L_{x}/2,L_{x}/2)\times(-L_{y}/2,L_{y}/2)$
with $L_{x}=12.8\,d_{i}$ and $L_{y}=25.6\,d_{i}$. Boundary conditions
are perfect conductor at $\pm L_{x}/2$, and periodic at $\pm L_{y}/2$.
The transport coefficients are: $\nu_{i}=5\times10^{-2}$, $\nu_{e}=10^{-4}$,
$\eta=5\times10^{-3}$, and $\kappa=2\times10^{-2}$. The specific
heat ratio is $\gamma=5/3$.

The simulation in both codes is started by perturbing $A_{z}$ with
a sinusoidal perturbation of the form:
\[
\delta A_{z}=-\epsilon\cos\left(\frac{\pi x}{L_{x}}\right)\cos\left(\frac{2\pi y}{L_{y}}\right),
\]
with $\epsilon=10^{-1}$. This problem was run with a 256$\times$256
packed mesh along both axes, with $\Delta x=2\times10^{-2}$ at $x=0$
(at the rational surface), and $\Delta y=5\times10^{-2}$ at $y=\pm L_{y}/2$
(the periodic boundaries). This represents a packing ratio vs. the
uniform mesh of 2.5 in $x$ and 2 in $y$. The time step is $\Delta t=0.05$.

In addition to the reconnected flux (which is a robust measure \citep{gem}),
we also track the temporal histories of kinetic ($\frac{1}{2}\int d\mathbf{x}\rho v^{2}$),
magnetic ($\frac{1}{2}\int d\mathbf{x}\,B^{2}$), and thermal ($\frac{1}{2}\int d\mathbf{x}\frac{p}{\gamma-1}$)
energies, which are more nuanced. The time history comparison is depicted
in Fig. \ref{fig:hifi_vs_pixie3d_gem}, and shows an early (slow)
phase followed by a late, explosive phase characterized by rapid growth
of kinetic energy and thermal energy, fed by a significant decrease
in magnetic energy. During this evolution, total energy is conserved
better than 0.04\%. The comparison shows remarkably good agreement
between HiFi and PIXIE3D during the slow phase and most of the explosive
phase, with some deviation (of about 1\%) late in the nonlinear evolution
of the problem. Key elements of comparison are the excellent agreement
in the timing of the transition from slow to explosive phase, in the
slopes during most of the explosive phase, and in the extrema values
of the energies at saturation. The reconnected flux for PIXIE3D is
also shown, along with data from Ref. \citep{gem} for the fully kinetic
particle simulation and the Hall MHD simulation. While details differ
in terms of timing and saturation level, the agreement in the slope
(maximum reconnection rate) between all three simulations is apparent.
The reconnected-flux history itself is not, but it is in very good
qualitative agreement with more recent HMHD GEM results, e.g. Fig.
13 of Ref. \citep{srinivasan2011analytical}.

\begin{figure}
\begin{centering}
\includegraphics[viewport=220bp 40bp 750bp 910bp,clip,width=0.3\columnwidth]{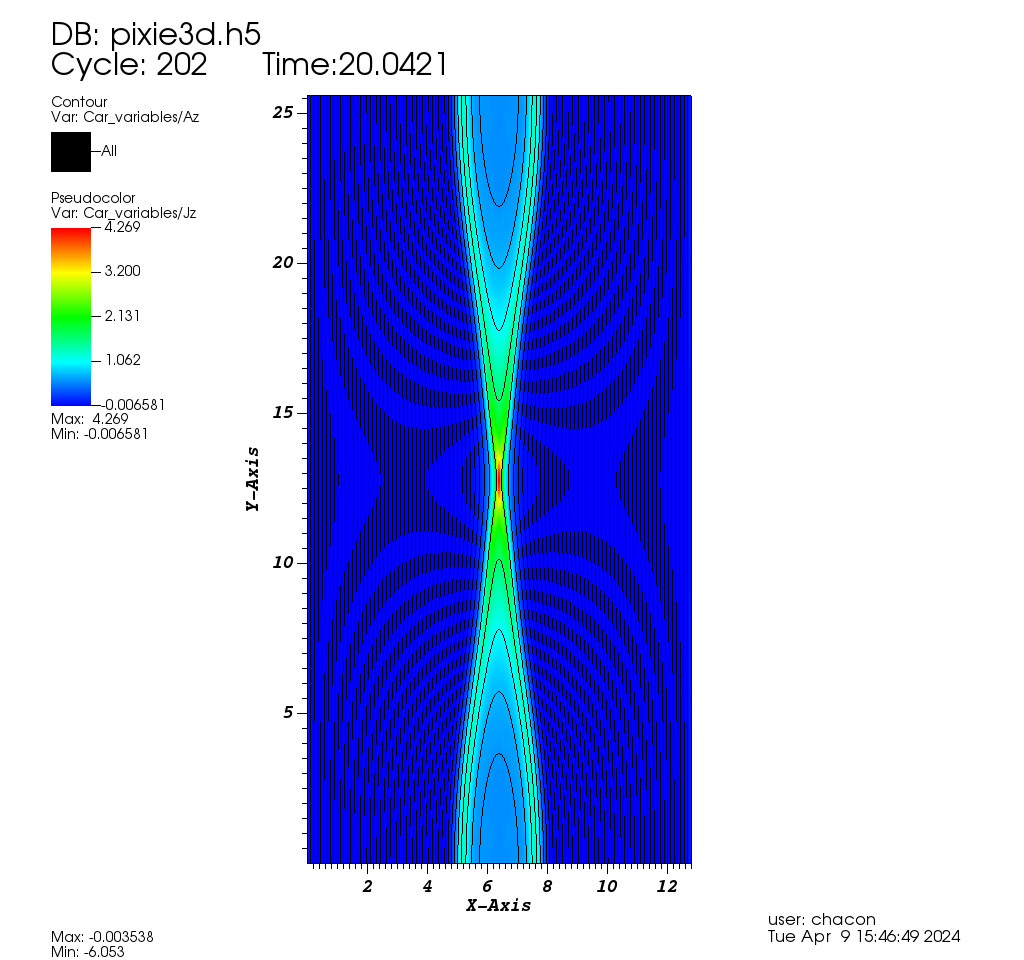}\includegraphics[viewport=220bp 40bp 750bp 910bp,clip,width=0.3\columnwidth]{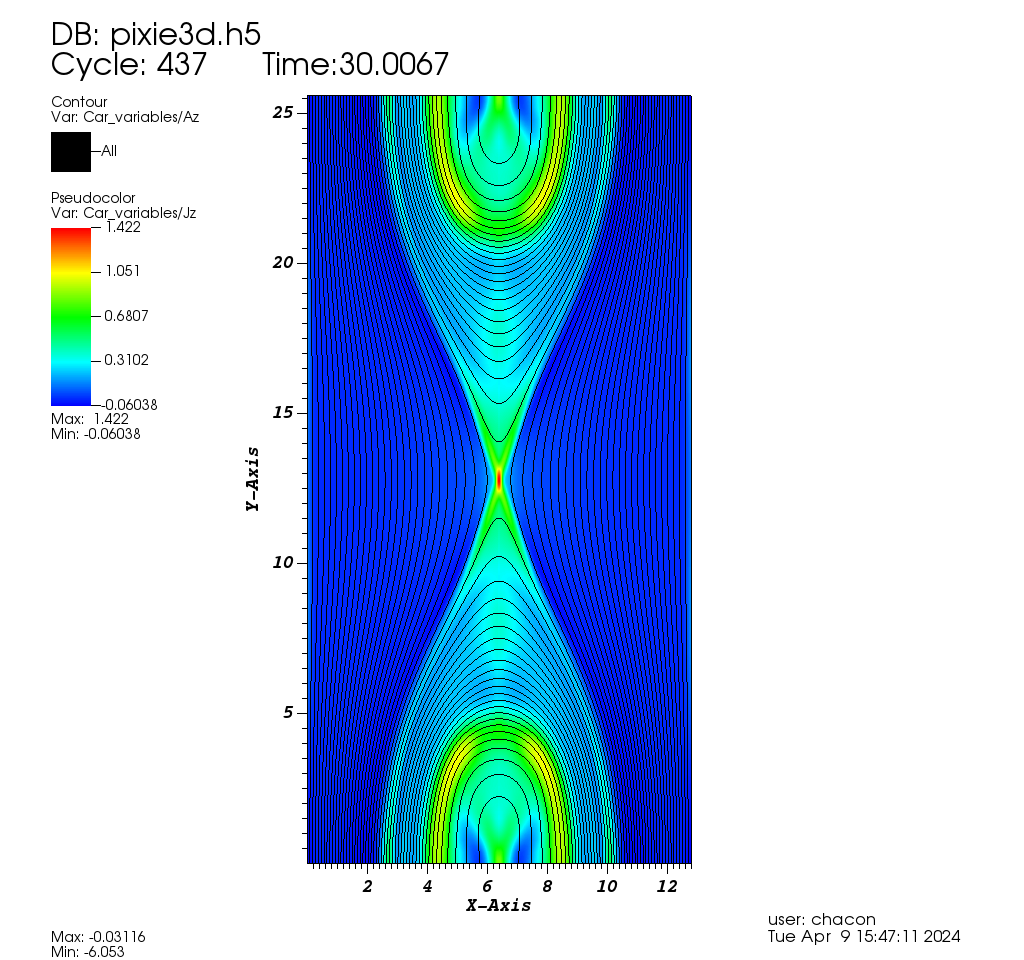}\includegraphics[viewport=220bp 40bp 750bp 910bp,clip,width=0.3\columnwidth]{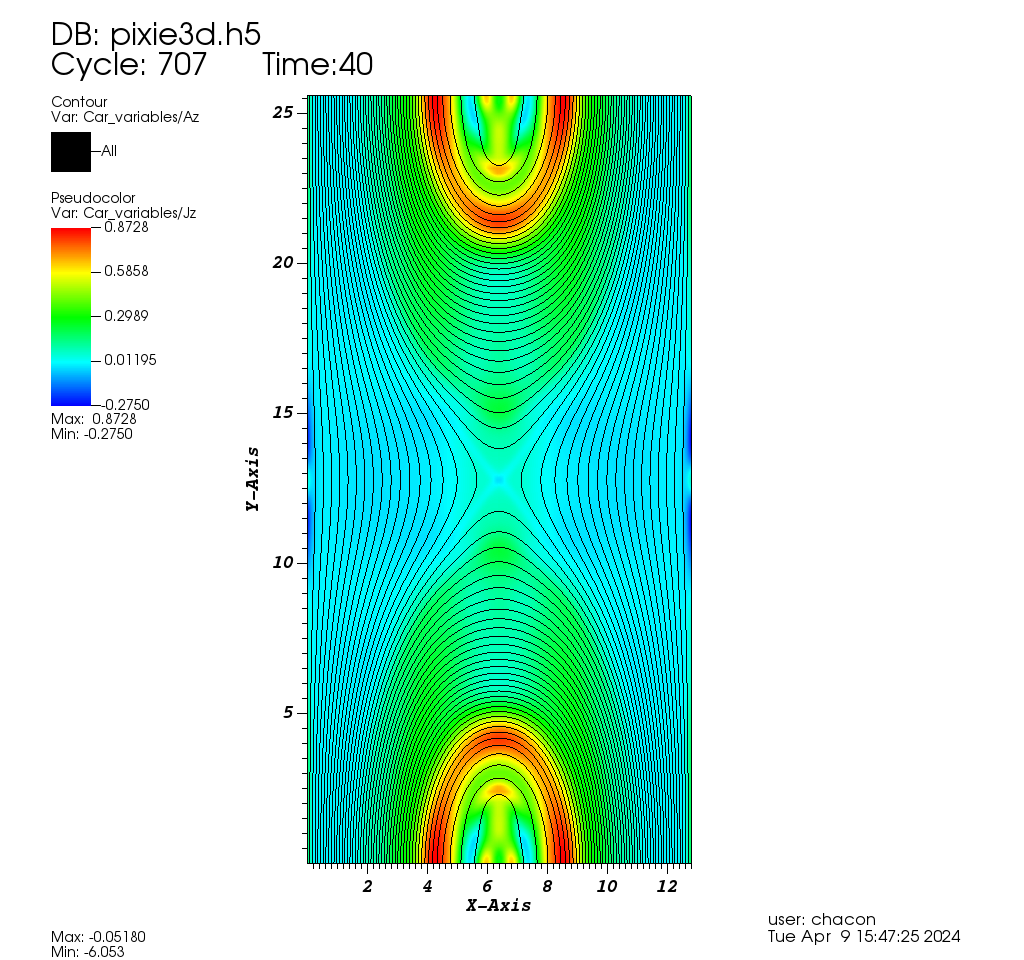}
\par\end{centering}
\caption{\label{fig:GEM-snapshots}Snapshots of $A_{z}$ (contours) and $j_{z}$
(color) at three different times ($t=20,30,40$) for the GEM simulation.}
\end{figure}
Snapshots of the $z$-component of the vector potential and the current
density for the GEM challenge at $t=20$ (before the explosive phase),
$t=30$ (at the peak of the explosive phase), and $t=40$ (at the
end of the simulation) are depicted in Fig. \ref{fig:GEM-snapshots}.
The formation of an open X-point and associated current density spike
is apparent. We do not find the formation of a central magnetic island
for our particular choice of parameters, which has been the subject
of contention in the literature \citep{toth-jcp-08-hall_amr,crockatt2022implicit}.

\subsection{Parallel and algorithmic performance}

\begin{table}
\centering{}\caption{\label{tab:Parallel-weak-scaling-performance-gem}Parallel weak-scaling
performance for the GEM Challenge problem with $\Delta t=0.01$ and
$\eta=\nu_{i}=\kappa=5\times10^{-4}$. Results are obtained in full
domain and a uniform Cartesian mesh, and averaged over 100 time steps
with a fixed timestep $\Delta t=0.01$. In this table, WCT stands
for wall-clock time. Speedup measures the speedup vs and explicit
method.}
{\small{}}%
\begin{tabular}{c|cccccccc}
{\small{}$n_{p}$} & {\small{}mesh} & {\small{}$\nu_{e}$} & {\small{}$\frac{\Delta t}{\Delta t_{exp}}$} & {\small{}$\textrm{GMRES}/\Delta t$} & {\small{}$\textrm{Newton}/\Delta t$} & {\small{}WCT(s)} & {\small{}WCT$/$GMRES} & {\small{}Speedup}\tabularnewline
\hline 
\hline 
{\small{}$1$} & {\small{}$32\times32$} & {\small{}$10^{-4}$} & {\small{}1.0} & {\small{}2.0} & {\small{}2.0} & {\small{}132} & {\small{}66} & {\small{}0.04}\tabularnewline
{\small{}$4$} & {\small{}$64\times64$} & {\small{}$2.5\times10^{-5}$} & {\small{}2.1} & {\small{}2.6} & {\small{}2.6} & {\small{}172} & {\small{}66} & {\small{}0.07}\tabularnewline
{\small{}$16$} & {\small{}$128\times128$} & {\small{}$6.3\times10^{-6}$} & {\small{}8.3} & {\small{}2.0} & {\small{}2.0} & {\small{}182} & {\small{}105} & {\small{}0.26}\tabularnewline
{\small{}$64$} & {\small{}$256\times256$} & {\small{}$1.6\times10^{-6}$} & {\small{}33} & {\small{}2.0} & {\small{}2.0} & {\small{}303} & {\small{}151} & {\small{}0.63}\tabularnewline
{\small{}$256$} & {\small{}$512\times512$} & {\small{}$3.9\times10^{-7}$} & {\small{}130} & {\small{}2.0} & {\small{}2.0} & {\small{}321} & {\small{}160} & {\small{}3.0}\tabularnewline
{\small{}$1024$} & {\small{}$1024\times1024$} & {\small{}$9.8\times10^{-8}$} & {\small{}520} & {\small{}2.0} & {\small{}2.0} & {\small{}509} & {\small{}203} & {\small{}8.5}\tabularnewline
{\small{}$4096$} & {\small{}$2048\times2048$} & {\small{}$2.4\times10^{-8}$} & {\small{}2100} & {\small{}2.0} & {\small{}2.0} & {\small{}512} & {\small{}256} & {\small{}38}\tabularnewline
{\small{}16384} & {\small{}$4096\times4096$} & {\small{}$6.0\times10^{-9}$} & {\small{}8400} & {\small{}3.0} & {\small{}3.0} & {\small{}1130} & {\small{}377} & {\small{}75}\tabularnewline
\hline 
\end{tabular}
\end{table}
We assess the (weak) parallel and algorithmic scalability of our solver
with the GEM challenge problem, with slightly modified parameters
as indicated below. Preconditioner setup and linear and nonlinear
convergence tolerances are set as stated earlier. For this exercise,
we consider uniform meshes in full domain starting at $32\times32$
with $n_{p}=1$ MPI task, up to $4096\times4096$ meshes with $n_{p}=16384$
tasks. We consider a fixed implicit timestep ($\Delta t=0.01$) for
all experiments, so the problem is becoming numerically stiffer as
the number of MPI tasks grows. In this way, when increasing the processor
count we are able to test the ability of the algorithm to \emph{both}
scale with mesh refinement and to scale in parallel. To maintain the
simulation in the stiff hyperbolic regime, we decrease the hyperresistivity
coefficient as we refine the mesh according to Eq. \ref{eq:hyperres-scaling}
to ensure that high-order dissipation is only provided for the grid-scale
components of the solution.

The results of the mesh/processor convergence study are shown in Table
\ref{tab:Parallel-weak-scaling-performance-gem}. Timings have been
obtained on the CPU partition of the Chicoma computer at Los Alamos
National Laboratory, featuring 1792 nodes, each with two Rome 64 core
2.6 GHz CPUs (also called the AMD EPYC 7H12) with 512 GB per node.
Several observations are in order. Firstly, as advanced earlier, the
fixed implicit timestep results in a stiffer numerical setup as we
refine the mesh, and indicated by the increase in $\Delta t/\Delta t_{exp}$
(with $\Delta t_{exp}$ the explicit stability-limit CFL timestep).
Next, the total number of GMRES iterations per timestep (GMRES/$\Delta t$,
which includes all Newton iterations) shows a very weak dependence
with mesh resolution (and number of MPI tasks), attesting to the effectiveness
of our preconditioning strategy. The performance of our MG parallel
implementation is demonstrated by the WCT/GMRES column, which measures
the time per GMRES iteration, which is dominated by the MG preconditioning
step and features a $log(n_{p})$ increase. The favorable algorithmic
(per the GMRES/$\Delta t$ column) and parallel (per the WCT/GMRES
column) scalings result in an overall remarkable scaling of the wall-clock
time WCT, which only increases by a factor of 8.6 when increasing
the problem size by a factor of 16384. It also results in a speedup
of 75 for the finest mesh considered vs. an explicit integrator, with
performance cross-over between implicit and explicit happening at
moderate meshes (between 256$\times$256 and 512$\times$512; the
cross-over happens at such fine meshes due to the fairly conservative
solver-parameter choices of the MG solver for the inversion of the
EMHD block, which makes the preconditioner quite expensive).

\section{Conclusions}

\label{sec:conclusions}

We have developed a parallel, scalable multigrid-based physics-based
preconditioning strategy for the Hall MHD equation system. As in earlier
MHD preconditioning studies, the physics-based preconditioning approach
exploits the natural parabolization of the hyperbolic couplings that
occurs when one discretizes them implicitly. The block structure of
the Hall MHD Jacobian spoils the ``arrow'' structure that was so
convenient in resistive MHD, but can still be formally Schur-factorized
in much the same way leading to a coupled system of equations for
the fluid velocity components. The Schur factorization is rendered
practical by 1) using the small-bulk-flow approximation (as in earlier
studies), and 2) exploiting the natural separation of stiff timescales
between ions and electrons in Hall MHD to segregate the electromagnetic
diagonal block solve.

The electromagnetic diagonal block corresponds to the so-called electron
MHD model (EMHD), and supports fast (quadratic) dispersive waves and
high-order dissipation (hyperresistivity), which is critical for nonlinear
stability. We invert this system robustly and scalably with MG-preconditioned
GMRES (nested GMRES solves in the preconditioning stage are allowed
by FGMRES in JFNK). Several strategies have helped in rendering the
EMHD block MG-friendly. Firstly, we have chosen to solve the vector-potential
formulation of Hall MHD (instead of the more standard magnetic field
one), which allowed us to reformulate the EMHD block into a set of
two coupled vectorial equations with physical interpretation (vector
potential and current density, respectively) and well-defined boundary
conditions. This coupled set of equations is Jacobi-smoothed in a
fully coupled manner. Secondly, we use a deferred-correction approach
in our MG smoothing strategy, whereby we approximate the $\nabla\times\nabla\times$
operator (which relates the current density and the vector potential)
as a vector Laplacian for the construction of the Jacobi block diagonals.
This strategy is demonstrated by analysis to produce a strongly smoothing
damped Jacobi iteration, and therefore a robust MG algorithm.

We have verified our implementation linearly with two wave propagation
tests, a whistler wave and a KAW, and nonlinearly with the GEM challenge
problem, by direct comparison with the HiFi spectral extended MHD
code. We have demonstrated the excellent scalability properties of
our preconditioner algorithmically (by demonstrating a bounded iteration
count with grid refinement) and in parallel (by demonstrating logarithmic
scaling of wall clock time with the number of MPI tasks in a weak
scaling sense with up to 16384 processors).

Future work will focus on extending the algorithm to singular coordinate
systems (with application to cylindrical and toroidal geometries,
of interest for fusion plasma simulation), and to include separate
ion and electron temperatures, anisotropic heat transport, and gyroviscosity.

\section*{Acknowledgments}

\noindent We acknowledge useful discussions with J. N. Shadid and
A. Stanier. We also acknowledge V. Lukin for providing the GEM challenge
simulation data for comparison with the HiFi code. This work was supported
by Triad National Security, LLC under contract 89233218CNA000001 and
DOE Office of Applied Scientific Computing Research (ASCR). The research
used computing resources provided by the Los Alamos National Laboratory
Institutional Computing Program.

\bibliographystyle{ieeetr}
\bibliography{general,numerical_MHD,numerics,MHD,tearing,ic,reconnection,adgrid}

\appendix

\section{Analysis of the smoothing property of the damped-Jacobi iteration
for the linearized EMHD system}

\label{sec:JB_analysis}

We specialize the linearized EMHD system in Eq. \ref{eq:aux_dj} for
a homogeneous uniform plasma in a periodic domain with a constant
magnetic field (and therefore no background current) and no electron
flow or electron inertia. The corresponding linear system reads:
\begin{eqnarray}
\delta\mathbf{A}+\Delta t\theta d_{i}\left(v_{A}\delta\mathbf{j}\times\mathbf{B_{0}}-d_{i}\nu_{e}\nabla^{2}\delta\mathbf{j}\right) & = & \mathbf{0},\nonumber \\
\delta\mathbf{j}-\nabla\times\nabla\times\delta\mathbf{A} & = & \mathbf{0}.\label{eq:aux_dj-homogeneous-1}
\end{eqnarray}
We have explicitly added the Alfven speed $v_{A}$ to keep track of
dimensionless numbers later. In matrix form, the system can be written
as:
\[
\left[\begin{array}{cc}
\mathbb{I} & \Delta t\theta d_{i}\left[v_{A}\mathbb{I}\times\mathbf{B}_{0}-d_{i}\nu_{e}\nabla^{2}\right]\\
-\nabla\times\nabla\times & \mathbb{I}
\end{array}\right]\left[\begin{array}{c}
\delta\mathbf{A}\\
\delta\mathbf{j}
\end{array}\right]=\left[\begin{array}{c}
0\\
0
\end{array}\right].
\]
Calling $\vec{\xi}=(\delta\mathbf{A},\delta\mathbf{j})^{T}$, we can
write the damped Jacobi iteration as:
\begin{equation}
\vec{\xi}^{l+1}=(\mathbb{I}-\sigma D^{-1}A)\vec{\xi}^{l},\label{eq:JB-iteration}
\end{equation}
where $A$ is the matrix above, $D$ is its diagonal (element- or
block-wise), and $\sigma\in(0,1]$ is the damping parameter. For simplicity,
we restrict the analysis below to 2D and consider a uniform mesh with
$\Delta x=\Delta y=\Delta$ as a sufficiently representative case
for the challenges at hand.

In this study, we consider the block diagonal matrix formed by the
diagonal coefficients of all vector components at a given cell, with
one important approximation: $-\nabla\times\nabla\times\approx\nabla^{2}$
in the current density equation. As we shall see, this approximation
delivers a convergent Jacobi iteration while avoiding the MG difficulties
originating in the $\nabla\times\nabla\times$ term without changing
the solution to the linear system upon convergence. There results:
\begin{equation}
D=\left[\begin{array}{cc}
\mathbb{I} & \Delta t\theta d_{i}\left[v_{A}\mathbb{I}\times\mathbf{B}_{0}+d_{i}\nu_{e}\frac{4}{\Delta^{2}}\mathbb{I}\right]\\
-\frac{4}{\Delta^{2}}\mathbb{I} & \mathbb{I}
\end{array}\right]=\left[\begin{array}{cc}
\mathbb{I} & U\\
L & \mathbb{I}
\end{array}\right],\label{eq:jb-diagonal}
\end{equation}
where:
\[
U=\Delta t\theta d_{i}\left[v_{A}\mathbb{I}\times\mathbf{B}_{0}+d_{i}\nu_{e}\frac{4}{\Delta^{2}}\mathbb{I}\right]\,\,,\,\,L=-\frac{4}{\Delta^{2}}\mathbb{I}.
\]
The inverse of the diagonal matrix $D$ is exactly given by:
\[
D^{-1}=\left[\begin{array}{cc}
\mathbb{I} & -U\\
0 & \mathbb{I}
\end{array}\right]\left[\begin{array}{cc}
\mathbb{I} & 0\\
0 & P^{-1}
\end{array}\right]\left[\begin{array}{cc}
\mathbb{I} & 0\\
-L & \mathbb{I}
\end{array}\right]=\left[\begin{array}{cc}
\mathbb{I}+UP^{-1}L & -UP^{-1}\\
-P^{-1}L & P^{-1}
\end{array}\right]=\left[\begin{array}{cc}
P^{-1} & -\frac{\Delta^{2}}{4}\left[\mathbb{I}-P^{-1}\right]\\
\frac{4}{\Delta^{2}}P^{-1} & P^{-1}
\end{array}\right],
\]
where:
\[
P=\mathbb{I}-LU=\alpha\mathbb{I}+\beta(\mathbb{I}\times\mathbf{B}_{0}),
\]
with:
\begin{equation}
\alpha=1+\frac{4\nu_{e}d_{i}}{v_{A}\Delta^{2}}\beta\,\,;\,\,\beta=\frac{4\Delta t\theta v_{A}d_{i}}{\Delta^{2}}.\label{eq:alpha_beta}
\end{equation}
Here, $\beta\gg1$ is the whistler CFL number (which ensures we are
in the stiff hyperbolic limit). The ratio $\alpha/\beta$ determines
the dissipation regime. If we choose $\nu_{e}$ such that dissipation
is provided at the grid scale, and have $\nu_{e}\sim v_{A}\Delta^{2}/d_{i}$
and therefore $\alpha\sim\beta\gg1$. Weak and strong hyperresistivity
regimes are then defined by $\alpha\ll\beta$ and $\alpha\gg\beta$,
respectively. The matrix $P$ has an analytical inverse, given by:
\[
P^{-1}=\frac{\alpha\mathbb{I}-\beta(\mathbb{I}\times\mathbf{B}_{0})+\frac{\beta^{2}}{\alpha}(\mathbb{I}\cdot\mathbf{B}_{0})\mathbf{B}_{0}}{\alpha^{2}+\beta^{2}}.
\]

To study the convergence properties of the Jacobi iteration, we perform
a Von-Neumann analysis with the ansatz:
\[
\vec{\xi}^{l}=\lambda^{l}e^{i\mathbf{k}\cdot\mathbf{x}}\vec{\xi}^{0},
\]
where $\lambda$ is the Jacobi iteration amplification factor. Stability
of the iteration demands $|\lambda|\leq1$. The smoothing property
requires $|\lambda|<1$ for the high-wavenumber modes in the system.
With this ansatz, we can diagonalize our differential operators as:
\begin{eqnarray*}
\nabla^{2} & \rightarrow & -\left(k_{x}^{2}\mathrm{sinc}^{2}\left(\frac{k_{x}\Delta}{2}\right)+k_{y}^{2}\mathrm{sinc}^{2}\left(\frac{k_{y}\Delta}{2}\right)\right)=-\kappa_{L}^{2},\\
\nabla & \rightarrow & i\left(k_{x}\mathrm{sinc}(k_{x}\Delta),k_{y}\mathrm{sinc}(k_{y}\Delta)\right)=i\boldsymbol{\kappa}_{G},\\
\nabla\times\nabla\times & \rightarrow & -\boldsymbol{\kappa}_{G}\times\boldsymbol{\kappa}_{G}\times,
\end{eqnarray*}
where $\mathrm{sinc(x)=\sin(x)/x}.$ The JB iteration in Eq. \ref{eq:JB-iteration}
then yields the equation for the amplification factor:
\begin{equation}
(1-\lambda)\left[\begin{array}{cc}
\mathbb{I} & 0\\
0 & \mathbb{I}
\end{array}\right]=\sigma\left[\begin{array}{cc}
\mathbb{I}+K\left[P^{-1}-\mathbb{I}\right] & \frac{\Delta^{2}}{4}\gamma P^{-1}\\
\frac{4}{\Delta^{2}}KP^{-1} & \mathbb{I}+\gamma P^{-1}
\end{array}\right],\label{eq:amp_factor-1}
\end{equation}
where:
\[
\gamma=\frac{d_{i}\nu_{e}}{v_{A}}\beta\left(\kappa_{L}^{2}-\frac{4}{\Delta^{2}}\right)=(\alpha-1)\left(\hat{\kappa}_{L}^{2}-1\right),
\]
and:
\begin{eqnarray*}
K & = & \left[\mathbb{I}+\hat{\boldsymbol{\kappa}}_{G}\times\hat{\boldsymbol{\kappa}}_{G}\times\mathbb{I}\right],
\end{eqnarray*}
with:
\begin{eqnarray}
\hat{\kappa}_{L}^{2} & = & \frac{\Delta^{2}\kappa_{L}^{2}}{4}=\left(\frac{\Delta^{2}k_{x}^{2}}{4}\mathrm{sinc}^{2}\left(\frac{k_{x}\Delta}{2}\right)+\frac{\Delta^{2}k_{y}^{2}}{4}\mathrm{sinc}^{2}\left(\frac{k_{y}\Delta}{2}\right)\right)=\sin^{2}\left(\frac{k_{x}\Delta}{2}\right)+\sin^{2}\left(\frac{k_{y}\Delta}{2}\right)\in(0,2],\label{eq:k_L}\\
\hat{\boldsymbol{\kappa}}_{G} & = & \frac{\Delta\boldsymbol{\kappa}_{G}}{2}=\frac{1}{2}\left(k_{x}\Delta\mathrm{sinc}(k_{x}\Delta),k_{y}\Delta\mathrm{sinc}(k_{y}\Delta)\right)=\left(\mathrm{sin}(k_{x}\Delta),\mathrm{sin}(k_{y}\Delta)\right).\nonumber 
\end{eqnarray}
Therefore, Eq. \ref{eq:amp_factor-1} yields:
\begin{equation}
\left[\begin{array}{cc}
(\lambda-1+\sigma)\mathbb{I}+\sigma K\left[P^{-1}-\mathbb{I}\right] & \frac{\sigma\Delta^{2}}{4}\gamma P^{-1}\\
\frac{4\sigma}{\Delta^{2}}KP^{-1} & (\lambda-1+\sigma)\mathbb{I}+\sigma\gamma P^{-1}
\end{array}\right]=0.\label{eq:amp_factor_final}
\end{equation}
Equation \ref{eq:amp_factor_final} is a 2$\times$2 block matrix
with commuting blocks. The solution for $\lambda$ is given by the
determinant of the matrix being zero. The determinant can be found
as (proof via a Schur factorization):
\[
\det\left[\left(\lambda^{*}\right)^{2}P+\lambda^{*}\sigma[\gamma\mathbb{I}+K(\mathbb{I}-P)]-\sigma^{2}\gamma K\right]=0.
\]
Restricting the analysis to the the stiff hyperbolic limit {[}$\alpha\gg1$,
which implies $(\mathbb{I}-P)\approx-P$ and $\gamma\approx\alpha\left(\hat{\kappa}_{L}^{2}-1\right)${]},
we find:
\begin{equation}
\det\left[\left(\lambda^{*}\right)^{2}\frac{P}{\alpha}+\lambda^{*}\sigma\left[\left(\hat{\kappa}_{L}^{2}-1\right)\mathbb{I}-K\frac{P}{\alpha}\right]-\sigma^{2}\left(\hat{\kappa}_{L}^{2}-1\right)K\right]=0.\label{eq:characteristic-poly}
\end{equation}
Here:
\[
\frac{P}{\alpha}=\mathbb{I}+\frac{\beta}{\alpha}(\mathbb{I}\times\mathbf{B}_{0}),
\]
and:
\[
\frac{\beta}{\alpha}\approx\frac{v_{A}\Delta^{2}}{4\nu_{e}d_{i}}.
\]
Equation \ref{eq:characteristic-poly} will be the starting point
to analyze three dissipation regimes of interest, all of them in the
strongly hyperbolic limit ($\beta\gg1$): the strong hyperresistivity
regime, the weak one, and the balanced regime with hyperresistivity
dissipation at the grid scale.

\paragraph{Strong hyperresistivity regime: $\beta\ll\alpha$}

In this case, we have:
\[
\left(\frac{P}{\alpha}\right)^{-1}=\alpha P^{-1}\approx\mathbb{I}-\frac{\beta}{\alpha}(\mathbb{I}\times\mathbf{B}_{0}).
\]
There results:
\[
\det\left[\left(\lambda^{*}\right)^{2}\mathbb{I}+\lambda^{*}\sigma\left[\left(\hat{\kappa}_{L}^{2}-1\right)\left(\mathbb{I}-\frac{\beta}{\alpha}(\mathbb{I}\times\mathbf{B}_{0})\right)-K\right]-\sigma^{2}\left(\hat{\kappa}_{L}^{2}-1\right)K\left(\mathbb{I}-\frac{\beta}{\alpha}(\mathbb{I}\times\mathbf{B}_{0})\right)\right]=0.
\]
Taking the $\beta/\alpha\rightarrow0$ limit, we have:
\[
\det\left[\left(\lambda^{*}\right)^{2}\mathbb{I}+\lambda^{*}\sigma\left[\left(\hat{\kappa}_{L}^{2}-1\right)\mathbb{I}-K\right]-\sigma^{2}\left(\hat{\kappa}_{L}^{2}-1\right)K\right]=0.
\]
This leads to a sixth-order characteristic polynomial, with six roots.
In vector form:
\begin{equation}
\left[\lambda^{*}+\sigma\left(\hat{\kappa}_{L}^{2}-1\right)\right]\left[\xi\mathbf{A}-K\mathbf{A}\right]=0,\label{eq:diff-dominated-case}
\end{equation}
with $\xi=\frac{\lambda^{*}}{\sigma}.$ This has the solution (with
multiplicity three, applying to all vector components):
\[
\myBox{\lambda_{L}=1-\sigma\hat{\kappa}_{L}^{2}}.
\]

Taking components of the remainder of Eq. \ref{eq:diff-dominated-case},
and noting that {[}assuming $B_{0}=1$ without loss of generality
and defining $A_{\parallel}=\mathbf{A}\cdot\mathbf{B}_{0}$, $A_{k}=\hat{\boldsymbol{\kappa}}_{G}\cdot\mathbf{A}$,
$A_{\times,\parallel}=\mathbf{B}_{0}\cdot(\hat{\boldsymbol{\kappa}}_{G}\times\mathbf{A})${]}:

\[
\hat{\boldsymbol{\kappa}}_{G}\cdot K\mathbf{A}=A_{k}\,\,;\,\,\hat{\boldsymbol{\kappa}}_{G}\times K\mathbf{A}=K(\hat{\boldsymbol{\kappa}}_{G}\times\mathbf{A})\,\,;\,\,\mathbf{B}_{0}\cdot K\mathbf{A}=(1-\hat{\kappa}_{G}^{2})A_{\parallel}+A_{k}\hat{\kappa}_{\parallel,G},
\]
we find:
\begin{itemize}
\item $\mathbf{B}_{0}\cdot[\cdots]\Rightarrow\xi A_{\parallel}-(1-\hat{\kappa}_{G}^{2})A_{\parallel}+A_{k}\hat{\kappa}_{\parallel,G}=0$,
\item $\hat{\boldsymbol{\kappa}}_{G}\cdot[\cdots]\Rightarrow\xi A_{k}-A_{k}=0\Rightarrow\xi=1\Rightarrow\myBox{\lambda=1}$,
\item $\mathbf{B}_{0}\cdot\left[\hat{\boldsymbol{\kappa}}_{G}\times[\cdots]\right]\Rightarrow\xi A_{\times,\parallel}-(1-\hat{\kappa}_{G}^{2})A_{\times,\parallel}=0\Rightarrow\myBox{\lambda_{G}=1-\sigma\hat{\kappa}_{G}^{2}}$.
\end{itemize}
The first equation with $A_{k}=0$ (from the second equation for $\lambda\neq1$)
also gives;
\[
\myBox{\lambda_{G}=1-\sigma\hat{\kappa}_{G}^{2}}.
\]
We have now found all six eigenvalues for the Jacobi amplification
factor, and they are all stable ($|\lambda|\leq1$). However, only
$\lambda_{L}$ allows for the smoothing property. In particular, from
the definition of $\hat{\kappa}_{L}^{2}$ in Eq. \ref{eq:k_L}, for
the Nyquist mode $k_{x}=k_{x}=\pi/\Delta$ we have $\hat{\kappa}_{L}^{2}=2$
and therefore $\lambda_{L}=1-2\sigma$, which implies strong damping
for a suitably chosen $\sigma\in(0,1]$ (e.g., $\sigma=0.7$) and
therefore a robust smoothing property for error components at the
grid scale. This is not the case for $\lambda_{G}$, since for the
Nyquist mode $\hat{\kappa}_{G}^{2}=0$ and therefore $\lambda_{G}=1$.
This difference in smoothing behavior stems from the fact that $\hat{\kappa}_{G}^{2}$
originates in the central difference discretization of a gradient,
whereas $\hat{\kappa}_{L}^{2}$ originates in the usual Laplacian
discretization. That said, for the strong hyperresistivity regime,
$\lambda_{L}$ has multiplicity three and is affecting all vector
components simultaneously, and is therefore providing strong smoothing
to the whole system.

\paragraph{Weak hyperresistivity regime: $\alpha\ll\beta$}

In this case, we have:
\[
\left(\frac{P}{\alpha}\right)^{-1}=\alpha P^{-1}=\frac{\alpha^{2}/\beta^{2}\mathbb{I}-\alpha/\beta(\mathbb{I}\times\mathbf{B}_{0})+(\mathbb{I}\cdot\mathbf{B}_{0})\mathbf{B}_{0}}{\alpha^{2}/\beta^{2}+1}\approx(\mathbb{I}\cdot\mathbf{B}_{0})\mathbf{B}_{0}-\frac{\alpha}{\beta}(\mathbb{I}\times\mathbf{B}_{0}).
\]
Taking the $\alpha/\beta\rightarrow0$ limit, there results:
\[
\det\left[\left(\lambda^{*}\right)^{2}\mathbb{I}+\lambda^{*}\sigma\left[\left(\hat{\kappa}_{L}^{2}-1\right)(\mathbb{I}\cdot\mathbf{B}_{0})\mathbf{B}_{0}-K\right]-\sigma^{2}\left(\hat{\kappa}_{L}^{2}-1\right)K(\mathbb{I}\cdot\mathbf{B}_{0})\mathbf{B}_{0}\right]=0.
\]
In vector form, the determinant equation can be written as:
\[
\left(\lambda^{*}\right)^{2}\mathbf{A}+\lambda^{*}\sigma\left[\left(\hat{\kappa}_{L}^{2}-1\right)A_{\parallel}\mathbf{B}_{0}-K\mathbf{A}\right]-\sigma^{2}\left(\hat{\kappa}_{L}^{2}-1\right)A_{\parallel}K\mathbf{B}_{0}=0.
\]
Taking components of this equation, we find:
\begin{itemize}
\item $\mathbf{B}_{0}\cdot[\cdots]\Rightarrow\left(\lambda^{*}\right)^{2}A_{\parallel}+\lambda^{*}\sigma\left[\left(\hat{\kappa}_{L}^{2}+\hat{\kappa}_{G}^{2}-2\right)A_{\parallel}-A_{k}\hat{\kappa}_{\parallel,G}\right]-\sigma^{2}\left(\hat{\kappa}_{L}^{2}-1\right)(1-\hat{\kappa}_{\perp,G}^{2})A_{\parallel}=0$,
or:
\[
\left(\left(\frac{\lambda^{*}}{\sigma}\right)^{2}+\frac{\lambda^{*}}{\sigma}\left(\hat{\kappa}_{L}^{2}+\hat{\kappa}_{G}^{2}-2\right)-\left(\hat{\kappa}_{L}^{2}-1\right)(1-\hat{\kappa}_{\perp,G}^{2})\right)A_{\parallel}-\frac{\lambda^{*}}{\sigma}\hat{\kappa}_{\parallel,G}A_{k}=0.
\]
\item $\hat{\boldsymbol{\kappa}}_{G}\cdot[\cdots]\Rightarrow\left(\lambda^{*}\right)^{2}A_{k}+\lambda^{*}\sigma\left[\left(\hat{\kappa}_{L}^{2}-1\right)A_{\parallel}\hat{\kappa}_{\parallel,G}-A_{k}\right]-\sigma^{2}\left(\hat{\kappa}_{L}^{2}-1\right)A_{\parallel}\hat{\kappa}_{\parallel,G}=0$
or:
\[
(\lambda^{*}-\sigma)\left[\frac{\lambda^{*}}{\sigma}A_{k}+\left(\hat{\kappa}_{L}^{2}-1\right)\hat{\kappa}_{\parallel,G}A_{\parallel}\right]=0\Rightarrow\myBox{\lambda=1}\,\,;\,\,\frac{\lambda^{*}}{\sigma}A_{k}+\left(\hat{\kappa}_{L}^{2}-1\right)\hat{\kappa}_{\parallel,G}A_{\parallel}=0.
\]
\item $\mathbf{B}_{0}\cdot\left[\hat{\boldsymbol{\kappa}}_{G}\times[\cdots]\right]\Rightarrow\left(\lambda^{*}\right)^{2}A_{\times,\parallel}-\lambda^{*}\sigma(1-\hat{\kappa}_{G}^{2})A_{\times,\parallel}=0\Rightarrow\myBox{\lambda=1-\sigma}\,\,;\,\,\myBox{\lambda_{G}=1-\sigma\hat{\kappa}_{G}^{2}}$.
\end{itemize}
Solving the first two equations, we find (in terms of $\xi=\lambda^{*}/\sigma$):
\[
\myBox{\lambda=1-\sigma}\,\,;\,\,\xi^{2}+\xi\left(\hat{\kappa}_{L}^{2}+\hat{\kappa}_{G}^{2}-2\right)+\left(\hat{\kappa}_{L}^{2}-1\right)(\hat{\kappa}_{G}^{2}-1)=0,
\]
gives:
\[
\xi=1-\hat{\kappa}_{L}^{2}\,\,;\,\,\xi=1-\hat{\kappa}_{G}^{2},
\]
which finally results in:
\[
\myBox{\lambda_{L}=1-\sigma\hat{\kappa}_{L}^{2}\,\,;\,\,\lambda_{G}=1-\sigma\hat{\kappa}_{G}^{2}}.
\]
For this case, we find that the problematic amplification factor $\lambda_{G}$
is present for all three components, but is always accompanied by
a strongly damping one, either $(1-\sigma)$ (for $A_{\times,\parallel}$)
or $\lambda_{L}$ (for $A_{\parallel}$, $A_{k}$), therefore resulting
in effective smoothing even without hyperresistivity.

\paragraph{Grid-scale hyperresistivity regime: $\beta/\alpha=1$}

For this case, we have:

\[
\det\left[\left(\lambda^{*}\right)^{2}\mathbb{I}+\lambda^{*}\sigma\left[\left(\hat{\kappa}_{L}^{2}-1\right)\alpha P^{-1}-K\right]-\sigma^{2}\left(\hat{\kappa}_{L}^{2}-1\right)K\alpha P^{-1}\right]=0
\]
with:
\[
\alpha P^{-1}=\frac{\mathbb{I}-(\mathbb{I}\times\mathbf{B}_{0})+(\mathbb{I}\cdot\mathbf{B}_{0})\mathbf{B}_{0}}{2}.
\]
This system can be expressed in vector form as:
\[
\left(\lambda^{*}\right)^{2}\mathbf{A}+\lambda^{*}\sigma\left[\left(\hat{\kappa}_{L}^{2}-1\right)(\alpha P^{-1})\mathbf{A}-K\mathbf{A}\right]-\sigma^{2}\left(\hat{\kappa}_{L}^{2}-1\right)K(\alpha P^{-1})\mathbf{A}=0,
\]
with:
\[
\alpha P^{-1}\mathbf{A}=\frac{\mathbf{A}-\mathbf{A}\times\mathbf{B}_{0}+A_{\parallel}\mathbf{B}_{0}}{2}.
\]
Taking components of this equation, noting that:
\begin{itemize}
\item $\hat{\boldsymbol{\kappa}}_{G}\cdot(\alpha P^{-1})\mathbf{A}=\frac{A_{k}-A_{\times,\parallel}+A_{\parallel}\hat{\kappa}_{\parallel,G}}{2}$.
\item $\mathbf{B}_{0}\cdot\left(\hat{\boldsymbol{\kappa}}_{G}\times(\alpha P^{-1})\mathbf{A}\right)=\frac{A_{\times,\parallel}-A_{\parallel}\hat{\kappa}_{\parallel,G}+A_{k}}{2}$.
\item $\mathbf{B}_{0}\cdot(\alpha P^{-1})\mathbf{A}=A_{\parallel}$.
\end{itemize}
we find:
\begin{itemize}
\item $\mathbf{B}_{0}\cdot[\cdots]\Rightarrow$\\
$\left(\left(\lambda^{*}\right)^{2}+\lambda^{*}\sigma\left(\hat{\kappa}_{G}^{2}+\hat{\kappa}_{L}^{2}-2\right)-\sigma^{2}\left(\hat{\kappa}_{L}^{2}-1\right)(1-\hat{\kappa}_{G}^{2})\right)A_{\parallel}-\lambda^{*}\sigma A_{k}\hat{\kappa}_{\parallel,G}-\sigma^{2}\left(\hat{\kappa}_{L}^{2}-1\right)\hat{\kappa}_{\parallel,G}\frac{A_{k}-A_{\times,\parallel}+A_{\parallel}\hat{\kappa}_{\parallel,G}}{2}=0,$
or:
\[
\left(\left(\frac{\lambda^{*}}{\sigma}\right)^{2}+\frac{\lambda^{*}}{\sigma}\left(\hat{\kappa}_{G}^{2}+\hat{\kappa}_{L}^{2}-2\right)-\left(\hat{\kappa}_{L}^{2}-1\right)(1-\hat{\kappa}_{G}^{2})\right)A_{\parallel}-\frac{\lambda^{*}}{\sigma}A_{k}\hat{\kappa}_{\parallel,G}-\left(\hat{\kappa}_{L}^{2}-1\right)\hat{\kappa}_{\parallel,G}\frac{A_{k}-A_{\times,\parallel}+A_{\parallel}\hat{\kappa}_{\parallel,G}}{2}=0
\]
\item $\hat{\boldsymbol{\kappa}}_{G}\cdot[\cdots]\Rightarrow\left(\left(\lambda^{*}\right)^{2}-\lambda^{*}\sigma\right)A_{k}+\left[\lambda^{*}\sigma-\sigma^{2}\right]\left(\hat{\kappa}_{L}^{2}-1\right)\frac{A_{k}-A_{\times,\parallel}+A_{\parallel}\hat{\kappa}_{\parallel,G}}{2}=0$,
or:
\[
(\lambda^{*}-\sigma)\left[\frac{\lambda^{*}}{\sigma}A_{k}+\left(\hat{\kappa}_{L}^{2}-1\right)\frac{A_{k}-A_{\times,\parallel}+A_{\parallel}\hat{\kappa}_{\parallel,G}}{2}\right]=0\Rightarrow\myBox{\lambda=1}\,\,;\,\,\frac{\lambda^{*}}{\sigma}A_{k}+\left(\hat{\kappa}_{L}^{2}-1\right)\frac{A_{k}-A_{\times,\parallel}+A_{\parallel}\hat{\kappa}_{\parallel,G}}{2}=0.
\]
\item $\mathbf{B}_{0}\cdot\left[\hat{\boldsymbol{\kappa}}_{G}\times[\cdots]\right]\Rightarrow$
\[
\left(\lambda^{*}\right)^{2}A_{\times,\parallel}+\lambda^{*}\sigma\left(\hat{\kappa}_{L}^{2}-1\right)\frac{A_{\times,\parallel}-A_{\parallel}\hat{\kappa}_{\parallel,G}+A_{k}}{2}-\lambda^{*}\sigma(1-\hat{\kappa}_{G}^{2})A_{\times,\parallel}-\sigma^{2}\left(\hat{\kappa}_{L}^{2}-1\right)(1-\hat{\kappa}_{G}^{2})\frac{A_{\times,\parallel}-A_{\parallel}\hat{\kappa}_{\parallel,G}+A_{k}}{2}=0,
\]
or:
\[
\left[\lambda^{*}-\sigma(1-\hat{\kappa}_{G}^{2})\right]\left[\lambda^{*}A_{\times,\parallel}+\sigma\left(\hat{\kappa}_{L}^{2}-1\right)\frac{A_{\times,\parallel}-A_{\parallel}\hat{\kappa}_{\parallel,G}+A_{k}}{2}\right]=0,
\]
and therefore:
\[
\myBox{\lambda_{G}=1-\sigma\hat{\kappa}_{G}^{2}}\,\,;\,\,\frac{\lambda^{*}}{\sigma}A_{\times,\parallel}+\left(\hat{\kappa}_{L}^{2}-1\right)\frac{A_{\times,\parallel}-A_{\parallel}\hat{\kappa}_{\parallel,G}+A_{k}}{2}=0.
\]
\end{itemize}
Solving the first two equations, we find:
\[
\left(\frac{\lambda^{*}}{\sigma}\right)^{2}+\frac{\lambda^{*}}{\sigma}\left(\hat{\kappa}_{G}^{2}+\hat{\kappa}_{L}^{2}-2\right)+\left(\hat{\kappa}_{L}^{2}-1\right)\left(\hat{\kappa}_{G}^{2}-1\right)=0,
\]
or:
\[
\myBox{\lambda_{L}=1-\sigma\hat{\kappa}_{L}^{2}\,\,;\,\,\lambda_{G}=1-\sigma\hat{\kappa}_{G}^{2}}.
\]
Else, $A_{\parallel}=0$, and from the last two equations we find:
\begin{eqnarray*}
\lambda^{*}A_{\times,\parallel} & + & \frac{\sigma\left(\hat{\kappa}_{L}^{2}-1\right)}{2}\left(A_{\times,\parallel}+A_{k}\right)=0,\\
\lambda^{*}A_{k} & + & \frac{\sigma\left(\hat{\kappa}_{L}^{2}-1\right)}{2}\left(A_{k}-A_{\times,\parallel}\right)=0.
\end{eqnarray*}
Solving, we find two complex conjugate roots:
\[
\myBox{\lambda=1-\sigma\left(1+\frac{\hat{\kappa}_{L}^{2}-1}{2}\right)\pm i\sigma\frac{|\hat{\kappa}_{L}^{2}-1|}{2}}.
\]
Recall that $|\hat{\kappa}_{L}^{2}-1|\in[0,1]$, that for stability
of the Jacobi iteration we need $|\lambda|\leq1$, and for the smoothing
property we need the strict inequality for the Nyquist mode. Computing
$|\lambda|$, we find:
\begin{equation}
|\lambda|^{2}=(1-\sigma)^{2}-(1-\sigma)\sigma(\hat{\kappa}_{L}^{2}-1)+\sigma^{2}\frac{(\hat{\kappa}_{L}^{2}-1)^{2}}{2},\label{eq:complex_roots}
\end{equation}
and therefore it follows that:
\[
|\lambda|<1\Rightarrow\sigma<\frac{2(1+\hat{\kappa}_{L}^{2})}{1+\hat{\kappa}_{L}^{4}}.
\]
This condition is always satisfied, as the right hand side is always
greater than unity for $\hat{\kappa}_{L}^{2}\in[0,2]$, and $\sigma\in(0,1]$.
Therefore, these roots are stable and strongly damping for the Nyquist
mode $\hat{\kappa}_{L}^{2}=2$. This can be shown by introducing $\hat{\kappa}_{L}^{2}$
into Eq. \ref{eq:complex_roots}, to find:
\[
|\lambda|^{2}=1-3\sigma+\frac{5}{2}\sigma^{2},
\]
which is always less than unity for $\sigma\in(0,1],$ and results
in very strong damping of the Nyquist mode for a suitably chosen damping
parameter: $|\lambda|^{2}=0.1$ for $\sigma=0.6$.

As in previous cases, we find the problematic amplification factor
$\lambda_{G}$ is present for all vector components, but always accompanied
by a robustly smoothing one: $\lambda_{L}$ for $A_{\parallel}$ and
two robustly smoothing complex conjugate amplification factors for
$A_{\times,\parallel}$ and $A_{k}$. These ensure the smoothing property
of the Jacobi iteration for the whole system.
\end{document}